\documentclass[10pt]{IEEEtran}
\IEEEoverridecommandlockouts
\usepackage[english]{babel}
\usepackage{cite,notoccite}
\usepackage{tikz,graphicx,subfigure,epstopdf}
\usepackage{lipsum,setspace,multirow,array}
\usepackage{amsmath,amsfonts,amssymb,amscd,amsthm,amstext}
\usepackage{mleftright}
\newcommand{\pren}[1]{\mleft(#1\mright)}

\newtheorem{thm}{Theorem}%[section]
\newtheorem{cor}{Corollary}%[section]
\newtheorem{lem}{Lemma}%[section]
\newtheorem{defn}{Definition}
\newtheorem*{defn*}{Definition}

\newtheorem{remark}{Remark}%[section]
\newtheorem*{remark*}{Remark}
\newtheorem*{example*}{Example}%[section]

\DeclareMathOperator*{\defeq}{\overset{{\rm def}}{=}}

\DeclareMathOperator*{\one}{\mathbf{1}}

\DeclareMathOperator*{\mcalr}{\mathcal{R}}
\renewcommand{\d}[1]{\ensuremath{\mathtt{d}{#1}}}
\newcommand{\D}{\ensuremath{\mathtt{d}}}

\newcommand{\tod}{\overset{d}{\to}}
\newcommand{\wtilde}[1]{\widetilde{#1}}

\newcommand{\Lc}{\mathcal{L}}

\newcommand{\Dc}{\mathcal{D}}
\newcommand{\Ac}{\mathcal{A}}
\newcommand{\R}{\mathbb{R}}
\newcommand{\Cb}{\mathbb{C}}
\newcommand{\Cc}{\mathcal{C}}
\newcommand{\Pp}{\mathcal{P}}

\newcommand{\E}{\mathbb{E}}
\newcommand{\Pb}{\mathbb{P}}

\newcommand{\bfx}{\mathbf{x}}
\newcommand{\bfy}{\mathbf{y}}

\newcommand{\tail}{\bar{F}}

\newcommand{\tx}{\textrm{tx}}
\newcommand{\rx}{\textrm{rx}}
\newcommand{\sinr}{\mathrm{SINR}}

\newcommand{\Composite}{\mathrm{Composite}}
\newcommand{\Pareto}{\mathrm{Pareto}}

\newcommand{\cm}{c.m.}
\newcommand{\TP}[1]{\ensuremath{\mathcal{TP}_{r}}}
\newcommand{\STP}[1]{\ensuremath{\mathcal{STP}_{r}}}

\newcommand{\fref}[1]{Figure~\ref{#1}}
\newcommand{\tref}[1]{Table~\ref{#1}}
\newcommand{\eref}[1]{(\ref{#1})}

\newcommand{\aref}[1]{Appendix~\ref{#1}}
\newcommand{\lemref}[1]{Lemma~\ref{#1}}
\newcommand{\thmref}[1]{Theorem~\ref{#1}}

\newcommand{\corref}[1]{Corollary~\ref{#1}}

\begin{document}

\title{Performance Limits of Network Densification}
%\author{\IEEEauthorblockN{Van Minh Nguyen}}

\author{Van~Minh~Nguyen,~\IEEEmembership{Member,~IEEE,} 
	and Marios~Kountouris,~\IEEEmembership{Senior Member,~IEEE,}% <-this % stops a space
	\thanks{Manuscript received December 1, 2016; revised March 1, 2017; accepted March 23, 2017. Part of this paper was presented at the IEEE International Conference on Communications, Kuala Lumpur, Malaysia, May 2016 \cite{Nguyen2016}.
	}
	\thanks{The authors are with the Mathematical and Algorithmic Sciences Lab, France Research Center, Huawei Technologies France SASU, 92100 Boulogne-Billancourt, France (email: vanminh.nguyen@huawei.com; marios.kountouris@huawei.com).}
	\thanks{Digital Object Identifier 10.1109/JSAC.2017.2687638}
	}

\markboth{IEEE JOURNAL ON SELECTED AREAS IN COMMUNICATIONS}{NGUYEN AND KOUNTOURIS: Performance Limits of Network Densification}

\maketitle

% ----------------------------------------------------------------
\begin{abstract}
	Network densification is a promising cellular deployment technique that leverages spatial reuse to enhance coverage and throughput. Recent work has identified that at some point ultra-densification will no longer be able to deliver significant throughput gains. In this paper, we provide a unified treatment of the performance limits of network densification. We develop a general framework, which incorporates multi-slope pathloss and the entire space of shadowing and small scale fading distributions, under strongest cell association in a Poisson field of interferers. First, our results show that there are three scaling regimes for the downlink signal-to-interference-plus-noise ratio (SINR), coverage probability, and average per-user rate. Specifically, depending on the near-field pathloss and the fading distribution, the user performance of 5G ultra dense networks (UDNs) would either monotonically increase, saturate, or decay with increasing network density. Second, we show that network performance in terms of coverage density and area spectral efficiency can scale with the network density better than the user performance does. Furthermore, we provide ordering results for both coverage and average rate as a means to qualitatively compare different transmission techniques that may exhibit the same performance scaling. Our results, which are verified by simulations, provide succinct insights and valuable design guidelines for the deployment of 5G UDNs.
\end{abstract}

\begin{IEEEkeywords}
	Cellular networks, 5G, network densification, extreme value theory, stochastic geometry, UDNs.
\end{IEEEkeywords}

% ----------------------------------------------------------------
\section{Introduction}\label{s:Introduction}
	Mobile traffic has significantly increased over the last decade mainly due to the stunning expansion of smart wireless devices and bandwidth-demanding applications. This trend is forecast to be maintained, especially with the deployment of 5G and beyond networks and machine-type communications. A major part of the mobile throughput growth during the past few years has been enabled by the so-called \emph{network densification}, i.e. adding more base stations (BSs) and access points and exploiting spatial reuse of the spectrum.
	Emerging fifth generation (5G) cellular network deployments are envisaged to be heterogeneous and dense, primarily through the provisioning of small cells such as picocells and femtocells. Ultra dense networks (UDNs) has been recognized as a promising solution to boost capacity and enhance coverage with low-cost and power-efficient infrastructure in 5G networks. It is advocated that UDNs will be the main technology enabler for achieving the 5G requirement of 1000x increase in mobile network data throughput compared to LTE. The underlying foundation of this claim is the presumed linear capacity scaling with the number of small cells deployed in the network. Moreover, recent advances on transport networks, such as high-capacity optical, millimeter wave (mmWave) communication and directional beamforming, may provide reliable and high-capacity links between the core network and small cells. 
	
	Nevertheless, there has been noticeable divergence between the above outlook and conclusion of various network studies according to which densification is not always beneficial to the network performance. Recent and often conflicting findings based on various modeling assumptions have identified that densification may eventually stop at a certain point delivering significant throughput gains. Using an unbounded pathloss model and Rayleigh fading, \cite{Andrews2011,Dhillon2012} show that the coverage probability does not depend on the network density when the background noise is negligible. By contrast, using a dual slope pathloss model, Rayleigh fading, and nearest BS association,  authors in \cite{Zhang2015} show that both coverage and capacity strongly depend on the network density. More precisely, coverage, expressed in terms of signal-to-interference-plus-noise ratio (SINR), is maximized at some finite density and there exists a phase transition in the near-field region with ultra-densification (i.e. network density goes to infinity). In \cite{Chen2012}, the authors consider strongest cell association with bounded pathloss and lognormal shadowing and show that the coverage attains a maximum point before starting to decay when the network becomes denser. In \cite{Arn16}, similar conclusions are obtained under Nakagami fading for the line-of-sight (LoS) and both nearest and strongest BS association. Based on system-level simulations and strongest cell association, \cite{LopezPerez2015} shows that there is an upper bound achieved at 1 cell per user, although such ultra dense deployments are neither cost nor energy efficient. Using multi-slope pathloss and smallest pathloss association, \cite{LopezPerez2016} shows that the network coverage probability first increases with BS density, and then decreases. The area spectral efficiency will grow almost linearly as the BS density goes asymptotically large. In \cite{Liu2016}, near-field communications is taken into account and shows that over-densification is harmful to the network performance. Optimal densification in terms of maximum SINR-coverage probability is investigated in \cite{Samarasinghe2014}. In \cite{Baccelli2015}, interference scaling limits in a Poisson field with singular power law pathloss and Rayleigh fading are derived. Moreover, authors in \cite{Lee2016} provide spectral efficiency scaling laws with spatial interference cancellation at the receiver. It is shown that linear scaling of the spectral efficiency with network density can be obtained if the number of receive antennas increases super-linearly with the network density (or linearly in case of bounded pathloss). 
	
	Wireless links are susceptible to time-varying channel impediments, interference and noise. It includes long-term attenuation due to pathloss, medium-term variation due to shadowing, and short-term fluctuations due to multi-path fading. In principle, these three main variations are usually taken into account by different network processing levels. Network planning usually depends on the effect of pathloss, while radio link-level procedures such as power control and handover aim at compensating for the combining effect of pathloss and shadowing, and fast fading is tackled by symbol-level processing at the physical layer. However, incorporating meaningfully the aforementioned propagation phenomena in network modeling often leads to complex and cumbersome mathematical derivations. As a result, most previous work, especially the ones using stochastic geometry, has nearly always assumed power law pathloss and Rayleigh fading due to their tractability. Although some works investigated the effect of pathloss singularity \cite{Inaltekin2009, Haenggi2009, Nguyen2011} or boundedness \cite{Zhang2015}, the effect of shadowing is usually ignored and other simplifications are employed for ease of analysis. This has led to unexpected observations in specific scenarios, as well as to divergent or even contrasting conclusions in prior work on the fundamental limits of network densification. Furthermore, the standard assumption of exponential distributed channel power will not apply in scenario with directional transmissions and multi-antenna processing, as are envisioned for massive MIMO, coordinated multipoint (CoMP) and mmWave systems. Advanced communication and signal processing techniques are expected to enhance the channel gain, which in some cases may have a diffuse power component \cite{Durgin2002} or regularly varying tail \cite{Rajan2015}. Therefore, it is relevant to analyze the performance of ultra dense networks considering general fading and shadowing.
	
	In this work, we aim at providing an answer to whether there are any fundamental limits to 5G UDNs due to physical limits arising from electromagnetic propagation. Specifically, we investigate the performance limits of network densification under a generic multi-slope pathloss model and general \emph{channel power} distribution, considering strongest cell association and Poisson distributed base stations. The term \emph{channel power} includes in this work all other propagation phenomena and transmission link gains except pathloss, including transmit power, small scale fading, shadowing, and gains due to antenna pattern, beamforming, etc. In particular, using general channel power distributions, we provide the ability to capture either the separate effect of small scale fading and of shadowing, or the combining effect of the composite fast fading-shadowing. Our general framework enables us to model any fading distribution and channel gain that can be observed in current wireless communications and networking. Furthermore, the generic multi-slope pathloss model not only covers both bounded and unbounded pathloss, but also captures the case of distance-dependent pathloss exponent. 
	Using tools from extreme value theory \cite{Embrechts1997}, and in particular regular variation analysis \cite{Bingham1989}, we propose a general framework and derive the scaling regimes of the downlink SINR, coverage probability, and average per-user rate. We first show the effect of both pathloss and channel power on the network performance with the following conclusions:
	\begin{itemize}
		\item Under the Poisson field assumption, the most affecting component of the pathloss is its near-field exponent $\beta_0$. Bounded pathloss (obtained for $\beta_0 = 0$ and widely used in the literature) is just a special case of $\beta_0 < d$ where $d$ is an integer denoting the network dimension. 
		\item The effect of channel power on the performance scaling is as significant as that of pathloss, and channel power following a regularly varying tail distribution has the same effect as the pathloss function's singularity.
	\end{itemize}
	Furthermore, we identify three fundamental scaling regimes for network densification:
	\begin{itemize}
		\item \emph{Growth regime}: Downlink SINR, coverage probability, and spectral efficiency always increase with the network density if the channel power distribution is \emph{slowly varying}, regardless of pathloss boundedness.
		\item \emph{Saturation regime}: All the above performance metrics saturate at a finite density point if either the near-field pathloss exponent is greater than the free-space dimension (i.e. $\beta_0 > d$), or channel power is \emph{regularly varying} with index in $(-1,0)$.
		\item \emph{Deficit regime}: The above performance metrics exhibit an `inverse U' behavior with respect to network density, i.e. they are maximized at a finite density then decay to zero when the network is further densified, if $\beta_0 < d$ and channel power follows any remaining tail than those in the above two regimes, i.e., any tail distribution that is not regularly varying with index in $[-1,0]$.
	\end{itemize}  
	
	Next, we investigate the scaling laws of network-level performance expressed in terms of \emph{coverage density} and \emph{area spectral efficiency}. We show that unlike per-user performance, network performance benefit more from the increasing network density. Most often, coverage density and area spectral efficiency scale linearly with the network infrastructure.
	
	Finally, using tools from stochastic orders \cite{Shaked1994}, we derive ordering results for both coverage probability and average rate in order to compare different transmission techniques and provide system design guidelines in general UDN settings, which may exhibit the same asymptotic performance. 
	
	The remainder is organized as follows. Section \ref{s:Model} describes the system model and defines the key performance metrics. Section \ref{sec:math} provides mathematical preliminaries. In Section \ref{sec:main}, we first develop scaling laws for SINR, coverage, and throughput for asymptotically large density and we identify three performance regimes for network densification; then we develop network performance limits and ordering. Finally, concluding remarks are given in Section \ref{s:Conclusion}.

\section{System Model}\label{s:Model}

\subsection{Network Model} 
	Consider a typical downlink user located at the origin and that the network is composed of cell sites located at positions $\{\bfx_i, i = 0,1,\ldots\}$. For convenience, cell sites are referred to as nodes, whereas the \emph{typical user} is simply referred to as \emph{user}. Unless otherwise stated, $\{\bfx_i\}$ are assumed to be random variables independently distributed on the \emph{network domain} according to a homogeneous Poisson point process (PPP) of intensity $\lambda$, denoted by $\Phi$. Users are distributed according to some independent and stationary point process $\Phi_u$ (e.g. PPP), whose intensity $\lambda_u$ is sufficiently larger than $\lambda$ in order to ensure that each BS is active, i.e. has at least one user associated within its coverage. In prior work, the entire $d$-dimensional Euclidean space $\R^d$ with $d = 2$ is usually assumed for network domain, where $\R$ is the set of real numbers. Since the network domain is in practice limited, the impact of far-away nodes is less relevant to the typical user due to pathloss attenuation, we assume that the distance from the user to any node is upper bounded by some constant $0 < R_{\infty} < \infty$. Each node transmits with some power that is independent to the others but is not necessarily constant. 

\subsection{Propagation Model}
\begin{figure}[!t]
		\centering
		\begin{tikzpicture}
		\clip (-3.6,-0.2) rectangle (3.6,3.6);
		\draw[thick] (0,0) circle (1.5cm); 
		\draw[thick,dashed] (0,0) circle (2.5cm);
		\draw[thick,dashed] (0,0) circle (3.5cm);
		% \draw[thick] (0,0) circle (6.5cm);
		\node[align=center] at (0,0.3) {$0$};
		\draw[thick,->] (0,0) -- node[near end,sloped,above=3pt]{$R_1$} (30:1.5);
		\draw[thick,->] (30:1.5) -- node[sloped,above=3pt]{$R_k$} (30:2.5);
		\draw[thick,->] (30:2.5) -- node[sloped,above=3pt]{$R_{k+1}$} (30:3.5);
		%		\draw[-{Stealth[scale=1.5]}] (0,0) -- node[near end,sloped,above=3pt]{$R_1$} (30:1.5);
		%		\draw[-{Stealth[scale=1.5]}] (30:1.5) -- node[sloped,above=3pt]{$R_k$} (30:2.5);
		%		\draw[-{Stealth[scale=1.5]}] (30:2.5) -- node[sloped,above=3pt]{$R_{k+1}$} (30:3.5);		
		% \draw[-{Stealth[scale=1.5]}] (30:5.0) -- node[near end,sloped,above=3pt]{$R_{\infty}$} (30:6.5);
		\node[align=center] at (-0.6,0.8) {$A_0 r^{\beta_0}$};
		\node[align=center] at (-2.2,2.0) {$A_k r^{\beta_k}$};
		\end{tikzpicture}
		\caption{Multi-slope pathloss model.}
		\label{fig:PLModel}
	\end{figure}
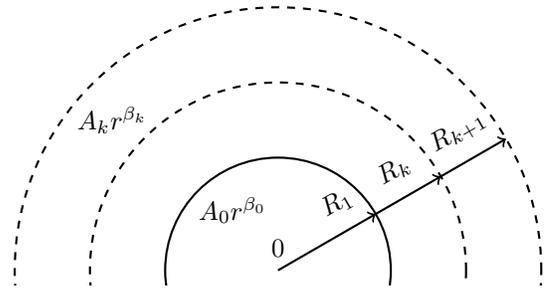

	Let $l: \R^+ \to \R^+$ represent the pathloss function, where $\R^+ = \{x \in \R | x \geq 0\}$. The receive power $P_{\rx}$ is related to the transmit power $P_{\tx}$ by $P_{\rx} = P_{\tx}/l(r)$ with $r$ being the transmitter-receiver distance. Physics laws impose that $1/l(r) \leq 1, \forall r$. However, in the literature, $l(\cdot)$ has been usually assumed to admit a singular power-law model, i.e. $l(r) \sim r^{\beta}$ where $\beta$ is the pathloss exponent satisfying $\beta \geq d$. This far-field propagation model has been widely used due to its tractability. However, for short ranges, especially when $r \to 0$, this model is no longer relevant and becomes singular at the origin. This is indeed more likely to happen in today's UDNs, where the inter-site distance becomes smaller. In addition, the dependence of the pathloss exponent on the distance in urban environments and in mmWave communications advocates the use of a more generic pathloss function. These requirements can be satisfied by modeling the pathloss function as follows (see \fref{fig:PLModel})
	\begin{equation}\label{eq:PL}
	    l(r) = \sum_{k=0}^{K-1} A_k r^{\beta_k} \one(R_k \leq r < R_{k+1}),
	\end{equation}
	where $\one(\cdot)$ is the indicator function, $K \geq 1$ is a given constant characterizing the number of pathloss slopes, and $R_k$ are constants satisfying
	\begin{equation}\label{eq:plRange}
	    %R_0 = 0, R_K = R_{\infty}, \text { and } R_k < R_{k+1}, \text{ for } k = 0,\ldots,K-1,
	    0 = R_0 < R_1 < \ldots < R_{K-1} < R_K = R_{\infty},
	\end{equation}
	$\beta_k$ denotes the pathloss exponent satisfying
	\begin{subequations}\label{eq:plExp}
	\begin{align}
	    \beta_0 & \geq 0, \label{eq:plExpA}\\
	    \beta_k & \geq d-1, \text{ for } k = 1,\ldots,K-1, \label{eq:plExpB}\\
	    \beta_k & < \beta_{k+1} < \infty, \text{ for } k = 0,\ldots,K-2, \label{eq:plExpC}
	\end{align}
	\end{subequations}
	and $A_k$ are constants to maintain continuity of $l(\cdot)$, i.e.
	\begin{equation}\label{eq:plScale}
	    A_k > 0, \text{ and } A_k R_{k+1}^{\beta_k} = A_{k+1} R_{k+1}^{\beta_{k+1}},
	\end{equation}
	for $k = 0,\ldots,K-2$. For notational simplicity, we also define $\delta_k = d/\beta_k$,  for $k = 0,\ldots,K-1$.
	
	The above general multi-slope model (see \eref{eq:PL}) captures the fact that the pathloss exponent varies with distance, while remaining unchanged within a certain range. In principle, free-space propagation in $\R^3$ has pathloss exponent equal to 2 (i.e. $\beta = d-1$), whereas in realistic scenarios, pathloss models often include antenna imperfections and empirical models usually result in the general condition \eref{eq:plExpB} for far-field propagation. Condition \eref{eq:plExpC} models the physical property that the pathloss increases faster as the distance increases. Notice, however, that this condition is not important in the subsequent analytical development. Finally, condition \eref{eq:plExpA} is related to the near field (i.e. it is applied to the distance range $[0, R_1]$). 
	
	% The multi-slope 
	The pathloss function as defined above has the following widely used special cases:
	\begin{itemize}
	    \item $K=1$, $\beta_0 \geq d$: $l(r) = A_0 r^{\beta_0}$, which is the standard singular (unbounded) pathloss model;
	    \item $K=2$, $\beta_0 = 0$: $l(r) = \max(A_0, A_1r^{\beta_1})$, which is the bounded pathloss recommended by 3GPP, in which $A_0$ is referred to as the minimum coupling loss.%, \cite{3GPP36931}.
	\end{itemize}
	
	Due to the particular importance of pathloss boundedness for having a realistic and practically relevant model, we have the following definition.
	\begin{defn}
	    A pathloss function $l: \R^+ \to \R^+$ is said to be \emph{bounded} if and only if $1/l(r) < \infty, \forall r \in \R^+$, and \emph{unbounded} otherwise. Furthermore, the pathloss function $l(\cdot)$ is said to be \emph{physical} if and only if  $1/l(r) \leq 1, \forall r \in \R^+$.
	\end{defn}
	It is clear that the pathloss function \eref{eq:PL} is bounded if and only if (iff) $\beta_0 = 0$, and is physical iff $\beta_0 = 0$ and $A_0 \geq 1$.
	
	Besides pathloss attenuation, shadowing and small-scale fading - commonly referred to as \emph{fading} in the sequel - are additional sources of wireless link variation. Let $m_i$ be a variable containing transmit power, fading, and any gain or attenuation other than pathloss from $i$-th node to the user, and let us refer to $m_i$ as \emph{channel power}. Then, given node location $\{\bfx_i\}$, the variables $\{m_i\}$ are assumed not identical to zero and are assumed independently distributed according to some distribution $F_m$. To this end, the signal power that the user receives from $i$-th node is expressed as $P_i = m_i / l(||\bfx_i||)$ where $||\cdot||$ is the Euclidean distance.
	
	From the above construction, $\{(\bfx_i, m_i)\}$ forms an independently marked (i.m.) Poisson point process, denoted by $\wtilde{\Phi}$. This allows to have the following definition.
	\begin{defn}
		With the above notation, a \emph{wireless network}, denoted by $\Xi$, is defined as the shot noise \cite{Baccelli2009} of $\wtilde{\Phi}$ associated with the response function defined as $v(\bfy,\bfx_i) := m_i/l(||\bfy - \bfx_i||)$, where $\bfy \in \mathbb{R}^d$ denotes the location of the receiver.
	\end{defn}
	
	\subsection{Performance Metrics}
	The signal quality experienced from $i$-th node, denoted by $Q_i$, is expressed in terms of its signal-to-interference-plus-noise ratio as
	\begin{equation*}
		Q_i = P_i / (I_i + W), %\frac{P_i}{I_i + W}
	\end{equation*}
	where $I_i = \sum_{j \in \Phi \setminus \{i\}} P_j$ is the aggregate interference experienced by the user served by the $i$-th node, and $W$ denotes the average power of the background thermal noise assumed to be Gaussian. In addition, we define 
	\begin{equation*}
		I = \sum_{j \in \Phi} P_j
	\end{equation*}
	to be the total interference. Note that $I = I_i + P_i$, $\forall i \in \Phi$.
	
	The performance metrics firstly studied in this paper are the coverage probability and the average rate that the user experiences from its \emph{serving cell}. Let $\sinr$ denote the signal quality that a user receives from its serving cell. The \emph{SINR coverage probability}, denoted by $\Pp_y$, is defined as the probability that $\sinr$ is larger than a given target value $y$, and the \emph{average rate}, denoted by $\Cc$, is defined as the Shannon rate (in nats/s/Hz) assuming Gaussian codebooks, i.e.
	\begin{equation}\label{eq:PerfDefn}
		\Pp_y = \Pb(\sinr \geq y), \text{ and } \Cc = \E(\log(1+\sinr)).
	\end{equation}
	
	Furthermore, we consider two system performance metrics, namely the \emph{coverage density} $\Dc_y$ (in BSs/m$^2$) and the {area spectral efficiency} (ASE) $\Ac$ (in nats/s/Hz/m$^2$), which are, respectively, defined as 
	\begin{equation}\label{eq:PerfDefn2}
		\Dc_y = \lambda\Pb(\sinr \geq y), \text{ and } \Ac = \lambda\E(\log(1+\sinr)).
	\end{equation}

    The coverage density gives an indication of the cell splitting gain \cite{Zhang2015}, i.e. the achievable data rate growth from adding more BSs due to the fact that each user shares its BS with a smaller number of users, as it provides the potential throughput if multiplied by the spectral efficiency $\log_2(1+y)$. % Interestingly, when $\lambda$ and $\lambda_u$ scale at the same rate, the scaling of ASE $\Ac$ and $\Dc_y$ will be the same. 

\subsection{User Association}
	The above performance metrics are defined with respect to the user's serving cell, which in turn depends on the underlying user association scheme. Nearest base station association has been widely employed in the literature of stochastic geometry based analyses mainly due to its mathematical convenience. In this work, we consider the realistic and practically relevant \emph{strongest cell association}, in which the user is connected to the cell site that provides the best signal quality or the strongest signal strength. Note that if $\sinr_{\max}$ and $\sinr_{\rm near}$ are the user's SINR under strongest cell association and nearest base station association, respectively, then by construction $\Pb(\sinr_{\rm near} \geq t) \leq \Pb(\sinr_{\max} \geq t), \forall t \in \R$. It can be easily shown that equality holds only if pathloss is not decreasing with respect to the distance and the channel power is a deterministic constant (implying that fading is absent or is not considered). In other words, nearest base station association is an under-approximation of the strongest cell association unless channel power is constant. Therefore, strongest cell association is the most appropriate scheme - both in a practical and theoretical sense - to assess the maximum achievable performance. In light of that, strongest cell association is considered in the sequel, and assuming each BS gives orthogonal resources (e.g. OFDMA) to users associated with it, the SINR of the typical user is given by
	\begin{equation}
		\sinr = \max_{i \in \Phi} Q_i,
	\end{equation} and can be expressed as \cite{Nguyen2010a,Nguyen2011}
	\begin{equation}\label{eq:YMI}
	    \sinr = \frac{M}{I + W - M}, \quad \text{with } M \defeq \max_{i \in \Phi}P_i.
	\end{equation}

\subsection{Notation} 
	Quantities whose dependence on the density $\lambda$ is analyzed, are denoted by $\cdot(\lambda)$, e.g. $\sinr(\lambda)$, $I(\lambda)$, and $M(\lambda)$.
	We also denote by $r$, $m$, and $P$, the distance, associated channel power, and received power from a random node, respectively. The distribution function of $P$ is denoted by $F_P$, and $\tail_P = 1 - F_P$. 
	
	We also use notation $\overset{d}{\to}$, $\overset{p}{\to}$, $\overset{a.s.}{\to}$ to denote the convergence in distribution, convergence in probability, and almost sure (a.s.) convergence, respectively. Notations $\Pb(\cdot)$ and $\E(\cdot)$ are respectively the probability and the expectation operators. 
	
	In addition, for real functions $f$ and $g$, %we say $f = O(g)$ if $\displaystyle \lim_{x \to \infty}(f(x)/g(x)) = c$ for $c \in (0, \infty)$, 
	we say $f \sim g$ if $\displaystyle \lim_{x \to \infty}(f(x)/g(x)) = 1$, and $f = o(g)$ if $\displaystyle \lim_{x \to \infty}(f(x)/g(x)) = 0$. 
	
	Finally, real intervals formed by numbers $a, b \in \R$ are denoted as follows: $(a,b) = \{x\in\R \, | \, a < x < b\}$, $[a,b) = \{x\in\R \, | \, a \leq x < b\}$, $(a,b] = \{x\in\R \, | \, a < x \leq b\}$, and $[a,b] = \{x\in\R \, | \, a \leq x \leq b\}$.

\section{Mathematical Preliminaries}\label{sec:math}

	Our characterization of the coverage and average rate scaling under general pathloss and fading models relies upon results from three related fields of study: regular variation, extreme value theory, and stochastic ordering.

\subsection{Regular Variation}
\begin{figure}[!t]
		\centering
		\begin{tikzpicture}[scale=0.75, every node/.style={scale=0.75}]
		%
		% Outer draw: light tails
		%
		\fill[gray] (0,0) ellipse (5cm and 3cm);
		\node[text width=5cm,align=left] at (0,2.3) {Light tails};	 
		%
		% Middle rectangle: Heavy tails
		%
		\draw[ultra thick,rounded corners,fill=white,rotate=0] (-3,-2) rectangle (3,1.5);
		\node[text width=6cm,align=center] at (0,0.7) {Heavy tails \\ 
			$\displaystyle \int_{\R}e^{\epsilon x} F(\d{x}) = \infty, \forall \epsilon > 0$} ;
		%
		% Inner eclipse: regular varying tails
		%
		\draw[ultra thick,rotate=15] (-0.5,-0.875) ellipse (2cm and 0.9cm) node[text width=3cm,align=center] {Regularly varying tails};
		\end{tikzpicture}
		\caption{Classification of tail behavior.}
		\label{fig:TailClass}
	\end{figure}
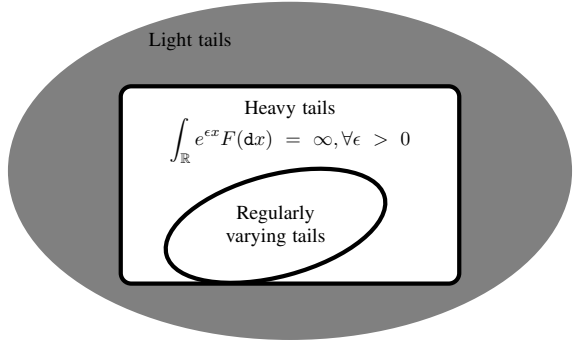

	\begin{defn}[Regular varying function in Karamata's sense \cite{Embrechts1997}] 
		A positive, Lebesgue measurable function $h$ on $(0,\infty)$ is called \emph{regularly varying} with index $\alpha \in \R$ at $\infty$ if $\displaystyle \lim_{x \to \infty} \frac{h(tx)}{h(x)} = t^{\alpha}$ for $0 < t < \infty$. In particular, $h$ is called \emph{slowly varying} (resp. \emph{rapidly varying}) (at $\infty$) if $\alpha = 0$ (resp. if $\alpha = -\infty$). We denote by $\mcalr_{\alpha}$ the class of regularly varying functions with index $\alpha$.
	\end{defn}
	Note that if $h$ is a regularly varying function with index $\alpha$, it can be represented as $h(x) = x^{\alpha} L(x)$ as $x \to \infty$ for some $L \in \mcalr_0$, see \cite[Theorem~1.4.1]{Bingham1989}. Moreover, if a function $h$ is regularly varying with index $\alpha$, $\alpha \in (-\infty,0]$, then it is heavy-tailed. In \fref{fig:TailClass}, we provide an illustration of the tail behavior classes; the interested reader is referred to \cite{Embrechts1997,Nguyen2011}.

	The following theorem due to Karamata is often useful to deal with regular varying functions.
	\begin{lem}[Karamata's theorem]\label{lem:KaramataThm}
		Let $L \in \mcalr_0$ be locally bounded in $[x_0,\infty)$ for some $x_0 \geq 0$. Then as $x \to \infty$,
		\begin{itemize}
			\item For $\alpha > -1$: $\displaystyle \int_{x_0}^x t^{\alpha}L(t)\d{t} \sim (\alpha+1)^{-1} x^{\alpha + 1} L(x)$;
	     	\item For $\alpha < -1$: $\displaystyle \int_{x}^{\infty} t^{\alpha}L(t)\d{t} \sim -(\alpha+1)^{-1}x^{\alpha+1}L(x)$.
		\end{itemize}
	\end{lem}
	
	For rapidly varying functions, we can have similar result as follows.
	\begin{lem}[Theorem A3.12, \cite{Embrechts1997}]\label{lem:RapidVary}
		Suppose $h \in \mcalr_{-\infty}$ is non-increasing, then for some $z > 0$ and all $\alpha \in \R$ we have: $\int_z^{\infty}t^{\alpha}h(t)\d{t} < \infty$ and
		\begin{equation}\label{eq:RapidVary}
			\lim_{x \to \infty}\frac{x^{\alpha+1}h(x)}{\int_{x}^{\infty}t^{\alpha}h(t)\d{t}} = \infty.
		\end{equation}
		Conversely, if for some $\alpha \in \R$, $\int_1^{\infty}t^{\alpha}h(t)\d{t} < \infty$ and \eref{eq:RapidVary} holds, then $h \in \mcalr_{-\infty}$.
	\end{lem}
		
	Another important theorem of Karamata theory is presented in the following.
	%	\begin{lem}[Monotone density theorem]\label{lem:MonotoneDensity}
	%		Let $U(x) = \int_0^x u(y)\d{y}$ (or $\int_x^{\infty} u(y)\d{y}$) where $u$ is ultimately monotone (i.e. $u$ is monotone on $(z,\infty)$ for some $z > 0$). If $U(x) \sim c x^{\alpha}L(x)$ as $x \to \infty$ with $c \geq 0$, $\alpha \in \R$ and $L \in \mcalr_0$, then $u(x) \sim c\alpha x^{\alpha-1}L(x)$ as $x \to \infty$. For $c = 0$ these relations are interpreted as $U(x) = o(x^{\alpha}L(x))$ and $u(x) = o(x^{\alpha-1}L(x))$.
	%	\end{lem}
	\begin{lem}[Monotone density theorem]\label{lem:MonotoneDensity}
		Let $U(x) = \int_0^x u(y)\d{y}$ (or $\int_x^{\infty} u(y)\d{y}$) where $u$ is ultimately monotone (i.e. $u$ is monotone on $(z,\infty)$ for some $z > 0$). 
		\begin{itemize}
			\item If $U(x) \sim c x^{\alpha}L(x)$ as $x \to \infty$ with $c \geq 0$, $\alpha \in \R$ and $L \in \mcalr_0$ (slowly varying), then $u(x) \sim c\alpha x^{\alpha-1}L(x)$ as $x \to \infty$. 
			\item For $c = 0$, the above relations are interpreted as $U(x) = o(x^{\alpha}L(x))$ and $u(x) = o(x^{\alpha-1}L(x))$.
		\end{itemize}					
	\end{lem}
	
	Finally, the following definition is useful for tail classification.	
	\begin{defn}[Tail-equivalence]
		Two distributions $F$ and $H$ are called \emph{tail-equivalent} if they have the same right endpoint, say $x_{\infty}$, and $\lim_{x \uparrow x_{\infty}} \bar{F}(x)/\bar{H}(x) = c$ for $0 < c < \infty$.
	\end{defn}

\subsection{Stochastic Ordering}
	Let $X$ and $Y$ be two random variables (RVs) defined on the same probability space such that $\mathbb{P}\left( X>t \right) \leq \mathbb{P}\left( Y>t \right), \forall t \in \mathbb{R}$.
	Then $X$ is said to be smaller than $Y$ in the \emph{usual stochastic order}, denoted by $X \leq_{\mathrm{st}} Y$, \cite{Shaked1994}. The interpretation is that $X$ is less likely than $Y$ to take on large values. 
	
	The above definition can be generalized for a set of real valued functions $g : (0, \infty) \to \mathbb{R}$ (denoted by $\mathcal{G}$), and $X$ and $Y$ be two non-negative random variables. The integral stochastic order with respect to $\mathcal{G}$ is defined as $X \leq_{\mathcal{G}} Y \Longleftrightarrow \mathbb{E}[g(X)] \leq \mathbb{E}[g(Y)], \forall g \in \mathcal{G}$.
	
	The following ordering using the Laplace transform is relevant to our paper. For the class $\mathcal{G} = \{g(x): g(x) = e^{-sx}, s > 0\}$, we have that 
	\begin{align}
		X \leq_{\mathrm{Lt}} Y \Longleftrightarrow \mathcal{L}_Y(s) & = \mathbb{E}[e^{-sY}] \nonumber\\
		& \leq \mathbb{E}[e^{-sX}] = \mathcal{L}_X(s), \forall s>0.\label{eqn:def_LT_ordering}
	\end{align}
	
	For all \emph{completely monotonic} (c.m.) functions $g(\cdot)$ (see the definition in the sequel), we have that $X \leq_{\mathrm{Lt}} Y \Longleftrightarrow \mathbb{E}[g(X)] \geq \mathbb{E}[g(Y)]$, whereas for all $g(\cdot)$ that have a completely monotonic derivative (c.m.d.), $X \leq_{\mathrm{Lt}} Y \Longleftrightarrow \mathbb{E}[g(X)] \leq \mathbb{E}[g(Y)]$.
		
	A function $g : (0,\infty) \rightarrow \mathbb{R}$ is said to be completely monotone ({\cm}), if it possesses derivatives of all orders which satisfy $(-1)^{n} g^{(n)}(x) \geq 0$, $\forall x \geq 0$ and $n \in \mathbb{N} \cup \lbrace 0 \rbrace$, where the derivative of order $n=0$ is defined as $g(x)$ itself. From Bernstein's theorem \cite{Bernstein1929}, a function is {\cm} iff it can be written as a mixture of decaying exponentials. 
	A function $g: (0,\infty) \rightarrow \mathbb{R}$ with $g(x) \geq 0, \forall x >0$, and
	% A function $g: (0,\infty) \rightarrow \mathbb{R}^+$ with 
	$\D g(x)/\D x$ being {\cm} is called a Bernstein function. Note that a {\cm} function is positive, decreasing and convex, while a Bernstein function is positive, increasing and concave. 
	
%%%%%%%%%%%%%%%%%%%%%%%%%%%%%%%%%%%%%%%%%%%%%%%%%%%%%%%%%%%%%%%%%%%%%%%%%%%%%%%%%%%
%%%%%%%%%%%%%%%%%%%%%%%%%%%%%%%%%%%%%%%%%%%%%%%%%%%%%%%%%%%%%%%%%%%%%%%%%%%%%%%%%%%
\section{Main Results}\label{sec:main}

\subsection{Problem Statement}
	The main objective of this work is to fully understand the potential, challenges and limitations of network densification, as well as to derive scaling laws for coverage and average rate. In particular, we aim at answering the following questions:
	\begin{itemize}
		\item \emph{How does network performance scale with the network infrastructure?}
		\item \emph{What should be done to leverage the potential of network densification?}
	\end{itemize}
	We investigate these questions from a technical point of view, using tools from regular variation, extreme value theory, and stochastic ordering. This framework, not only allows us to characterize the asymptotic performance under general fading and pathloss models, but it also provides useful system design guidelines even in the absence of analytical closed-form expressions. 

\subsection{Tail Behavior of Wireless Link}
	The network performance precisely depends on the received SINR, which in turn depends on $M$ and $I$ (see \eref{eq:YMI}). The behavior of the maximum $M$ and the sum $I$ is totally determined by that of the received power $P_i$. The following result characterizes the signal power $P_i$, in particular its tail behavior using tools from regular variation theory.

	\begin{thm}\label{thm:tailequiv} The tail distribution of the received signal power $\tail_P$ depends on the tail distribution of the channel power  $\tail_m$ and the pathloss function $l(r)$ as follows:
	\begin{itemize}
		\item If $\tail_m \in \mcalr_{-\alpha}$ with $\alpha \in [0,\infty]$, then $\tail_P \in \mcalr_{-\rho}$ where $\rho = \min(\delta_0, \alpha)$ with the convention that $\delta_0 = +\infty$ for $\beta_0 = 0$, and $\min(+\infty,+\infty) = +\infty$;
		\item If $\tail_m(x) = o(\bar{H}(x))$ as $x \to \infty$ with $\bar{H} \in \mcalr_{-\infty}$, then $\tail_P(t)$ and $\tail_m(A_0 t)$ are tail-equivalent for $\beta_0 =0$, and $\tail_P \in \mcalr_{-\delta_0}$ for $\beta_0 > 0$.
	\end{itemize}
	\end{thm}
	\begin{IEEEproof}
		\textnormal{See Appendix \ref{appen_Theorem1}.}
	\end{IEEEproof}
	
	\thmref{thm:tailequiv} shows that the tail behavior of the wireless link depends not only on whether the pathloss function is bounded or not, but also on the tail behavior of the channel power. More precisely, a key implication of \thmref{thm:tailequiv} is that pathloss and channel power have interchangeable effects on the tail behavior of the wireless link. This can also be shown using Breiman's Theorem \cite{Breiman1965} and results from large deviation of product distributions. Specifically, if the channel power is regularly varying with index $-\rho$ and the $\rho$-moment of the pathloss is finite, then the tail behavior of the received signal is governed by the regularly varying tail, i.e. the wireless link is also regularly varying regardless of the pathloss singularity. For lighter-tailed channel power, the regular variation property of the wireless link is solely imposed by the pathloss singularity.
	
	More importantly, \thmref{thm:tailequiv} is a general result and covers all tail behaviors of the channel power. The first case covers the heaviest tails (i.e. $\mcalr_{-\alpha}$ with $0 \leq \alpha < \infty$), such as Pareto distribution, as well as the moderately heavy tails (i.e. the class $\mcalr_{-\infty}$), such as exponential, normal, lognormal, gamma distributions. The second case covers all remaining tails (e.g. truncated distributions). Therefore, for any statistical distribution of the channel power, and in particular of fast fading and shadowing, \thmref{thm:tailequiv} allows us to characterize the tail behavior of the wireless link, which is essential to understand the behavior of the interference, the maximum received power, and their asymptotic relationship. 
		
	In current wireless networks, the signal distribution $\tail_m$ is governed by lognormal or gamma shadowing and Rayleigh fast fading. Since these fading distributions belong to the class $\mcalr_{-\infty}$ and the pathloss is bounded, the tail distribution of the received signal follows $\tail_P \in \mcalr_{-\infty}$. Note also that in most relevant cases, it can be shown that $\tail_P$ belongs to the maximum domain of attraction of a Gumbel distribution \cite{Nguyen2010a}.
	
	\begin{remark}
		The above result can be seen as a generalization of prior results: \cite{Haenggi2009} showed that the interference is tail-equivalent with the channel power if pathloss is bounded and if $\E(m) < \infty$ (note that this condition does not hold for $\tail_m \in \mcalr_{-\alpha}$, $\alpha \in [0,1]$). \cite[Chap.~8]{Nguyen2011} showed that under lognormal channel power, $\tail_P$ is regularly varying for unbounded pathloss, and behaves like a lognormal tail for bounded pathloss.
	\end{remark}
	
	Based on the above result, we want now to better understand the signal power scaling and the interplay between pathloss function and channel power distribution. The following result can be derived.
	
	\begin{cor} The tail distribution $\tail_P$ is classified as follows:
		\begin{itemize}
			\item $\tail_P \in \mcalr_0$ if and only if $\tail_m \in \mcalr_0$;
			\item $\tail_P \in \mcalr_{-\alpha}$ with $\alpha \in (0,1)$ if $\beta_0 > d$ or $\tail_m \in \mcalr_{-\alpha}$;
			\item $\tail_P \in \mcalr_{-\alpha}$ with $\alpha > 1$ if $0 < \beta_0 < d$ and $\tail_m \notin \mcalr_{-\rho}$ with $\rho \in [0,1]$;
			\item $\tail_P = o(\bar{H})$ with $\bar{H} \in \mcalr_{-\infty}$ if $\beta_0 = 0$ and $\tail_m = o(\bar{H})$.
		\end{itemize}
		\label{cor:classifyTail}
	\end{cor}
	
\subsection{SINR Scaling Laws}
	The tail behavior of the wireless link as characterized in the previous section allows to provide scaling laws for the received SINR in the high density regime.
	\begin{thm}\label{thm:Yinfty}
		In dense networks (i.e., $\lambda \to \infty$), the received SINR behaves as
		\begin{enumerate}
			\item $\sinr \overset{p}{\to} \infty$ if $\tail_P \in \mcalr_0$;
			\item $\sinr \overset{d}{\to} D$ if $\tail_P \in \mcalr_{-\alpha}$ with $0 < \alpha < 1$, where $D$ has a non-degenerate distribution;
			\item $\sinr \overset{a.s.}{\to} 0$ if $\tail_P \notin \mcalr_{-\alpha}$ with $\alpha \in [0,1]$.
		\end{enumerate}		
	\end{thm}
	\begin{IEEEproof}
		\textnormal{See Appendix \ref{appen_Theorem2}.}
	\end{IEEEproof}
	
	According to \corref{cor:classifyTail}, $\tail_P \in \mcalr_0$ is due to the fact that $\tail_m \in \mcalr_0$. Since $m$ is the channel power containing the transmit power and all potential gains and propagation phenomena (including fading, array gain, etc.), $\tail_m \in \mcalr_0$ means that the channel power is more probable to take large values. As a result, it compensates the pathloss and makes the desired signal power grow at the same rate as the aggregate interference (i.e. $M/I \overset{p}{\to} 1$). This provides a theoretical justification to the fact that network densification always enhances the signal quality $\sinr$.
	
\begin{figure}[!t]
		\centering
		\subfigure[$\beta_0 = 3$, $F_m \sim \Composite$]
		{
			\includegraphics[width=0.475\textwidth]{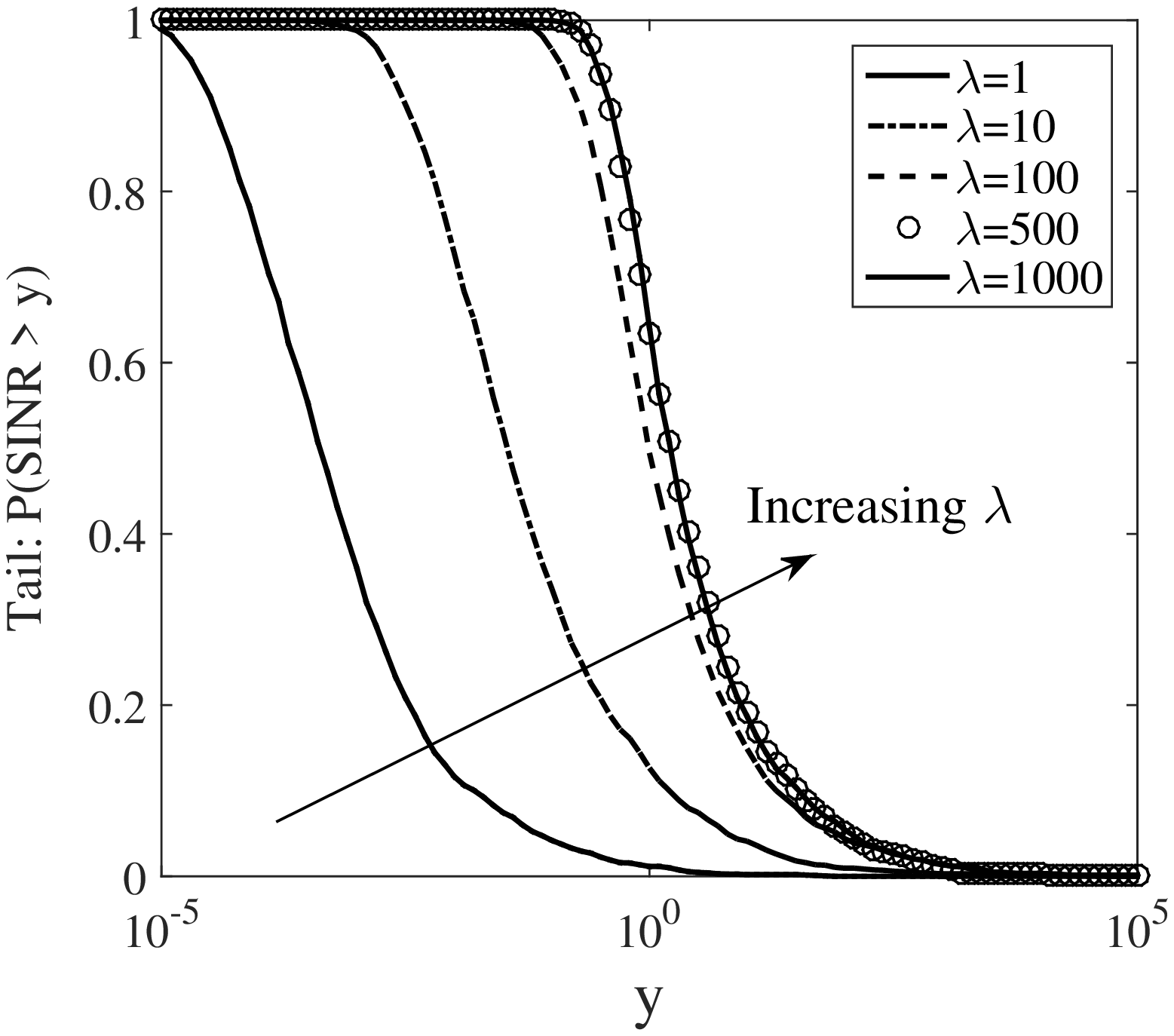}
		}
		\subfigure[$\beta_0 = 0$, $F_m \sim \Pareto(0.5)$]
		{
			\includegraphics[width=0.475\textwidth]{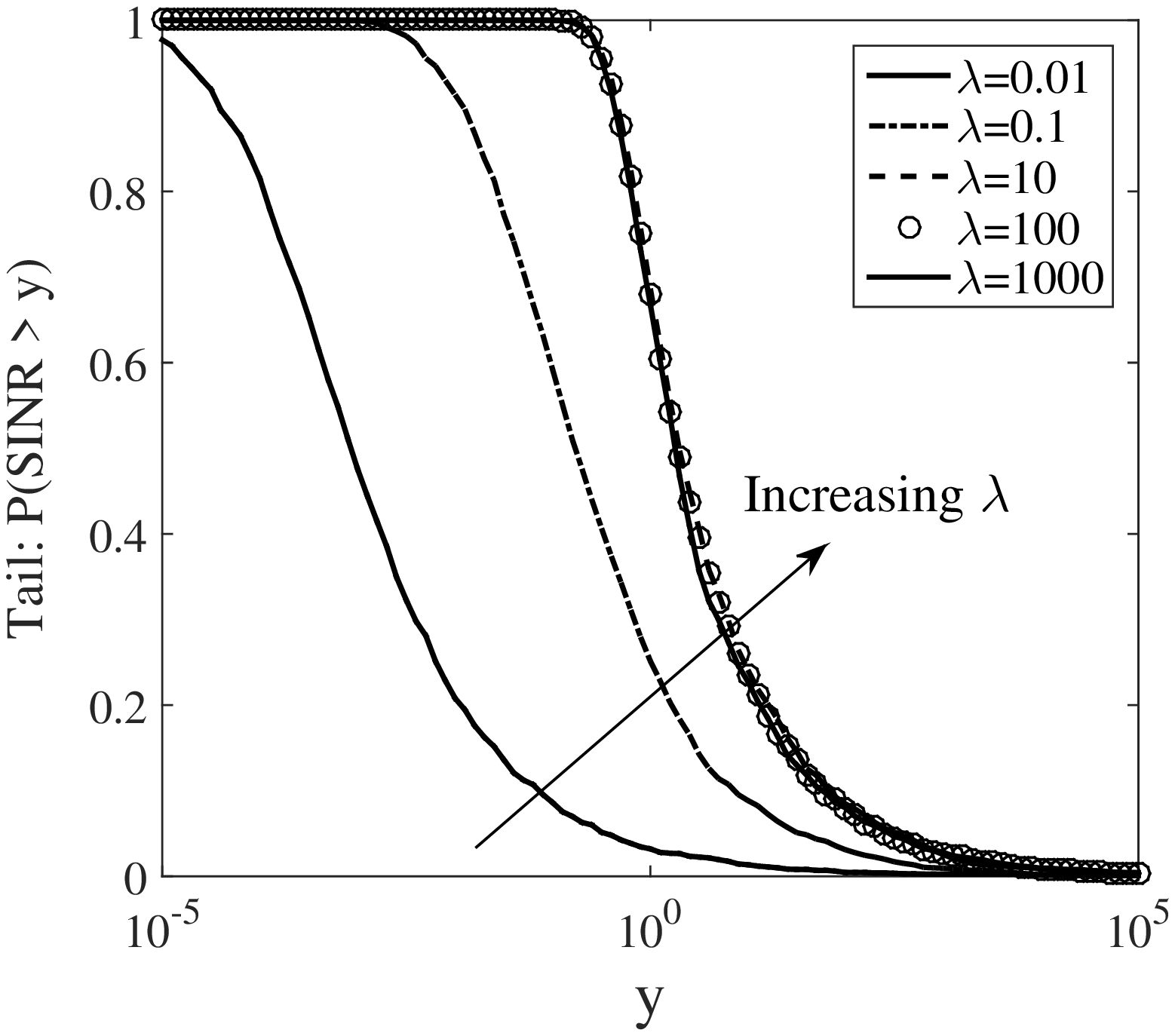}
		}
		\caption{Validation of the convergence of SINR to a steady distribution in case of $\tail_P \in \mcalr_{-\alpha}$ with $\alpha \in (0,1)$. Simulated parameters include two-slope pathloss (i.e. $K =2$) with $A_0 = 1$, $\beta_1 = 4$, $R_1 = 10$ m, $R_{\infty} = 40$ km, two-dimensional network domain (i.e. $d=2$), and network density $\lambda$ in BSs/$\text{km}^2$.}
		\label{fig:ConvergYtoD}
	\end{figure}

	When $\tail_P \in \mcalr_{-\alpha}$ with $0 < \alpha < 1$, $\sinr \overset{d}{\to} D$ implies that the SINR distribution converges to a non-degenerate distribution. Moreover, from \corref{cor:classifyTail}, this is due to either large near-field exponent or heavy-tailed channel power. In that case, for any SINR target $y$, the coverage probability $\Pb(\sinr > y)$ flattens out starting from some network density $\lambda$ (ceiling effect). This means that further increasing the network density by installing more BSs does not improve the network performance. This saturation effect is confirmed by simulation experiments shown in \fref{fig:ConvergYtoD}, where the tail distribution of $\sinr$ converges to a steady distribution for both cases: either $\beta_0 > d$ (left plot) or $\tail_m \in \mcalr_{-\alpha}$ with $\alpha \in (0,1)$ (right plot). In \fref{fig:ConvergYtoD}, by $F_m \sim \Composite$ we mean that channel power corresponds to the case with constant transmit power and with commonly known composite Rayleigh-lognormal fading, which belongs to the rapidly varying class $\mcalr_{-\infty}$. $\Pareto(\alpha)$ stands for channel power following a Pareto distribution of shape $1/\alpha$ and some scale $\sigma > 0$, i.e. $\Pareto(\alpha): \tail_m(x) = (1 + x/\sigma)^{-\alpha}$. Note that $\Pareto(\alpha) \in \mcalr_{-\alpha}$.
	
	\begin{figure}[!t]
		\centering
		\subfigure[$\beta_0 = 1$, $F_m \sim \Composite$]
		{
			\includegraphics[width=0.475\textwidth]{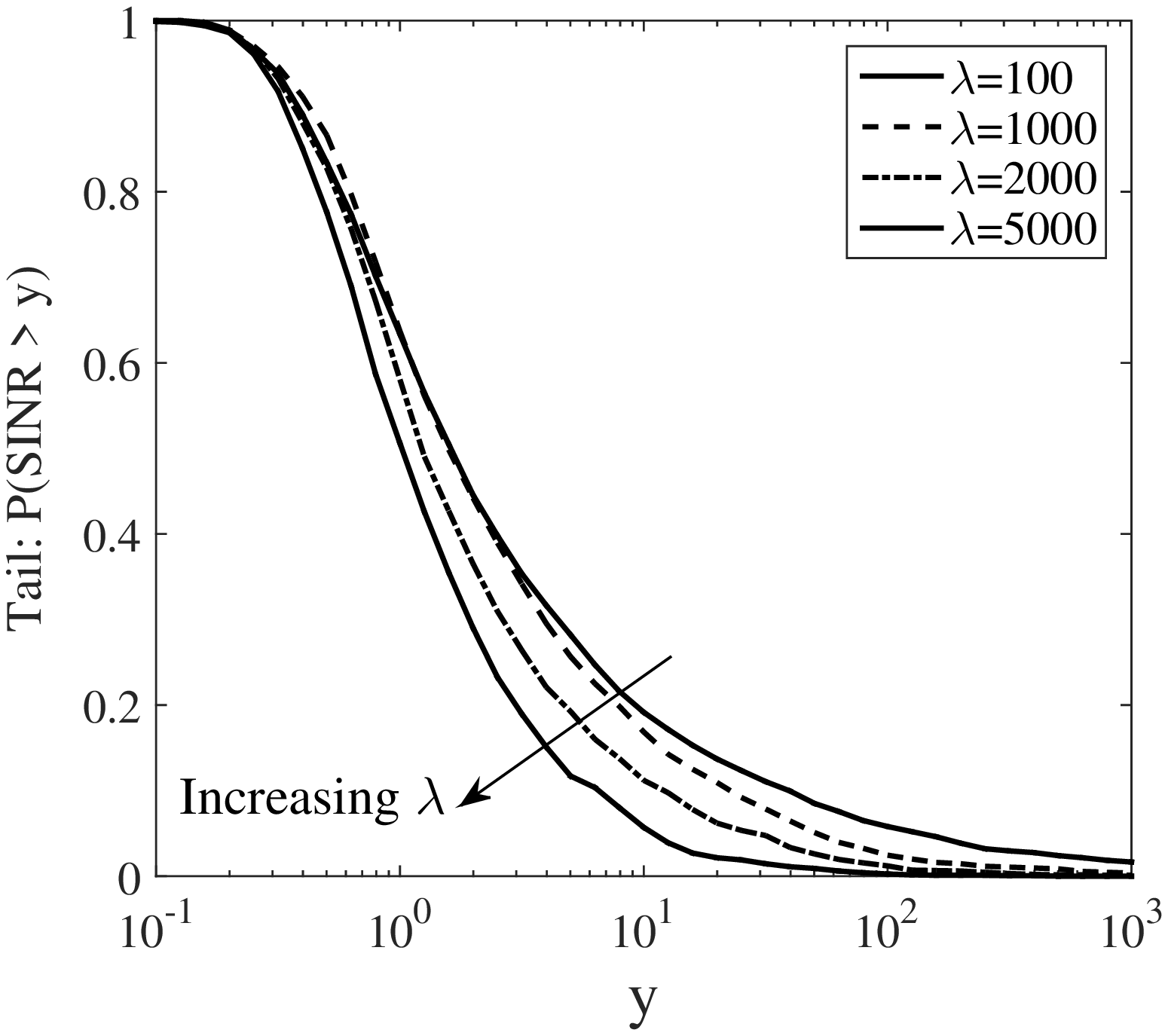}
		}
		\subfigure[$\beta_0 = 0$, $F_m \sim \Pareto(4)$]
		{
			\includegraphics[width=0.475\textwidth]{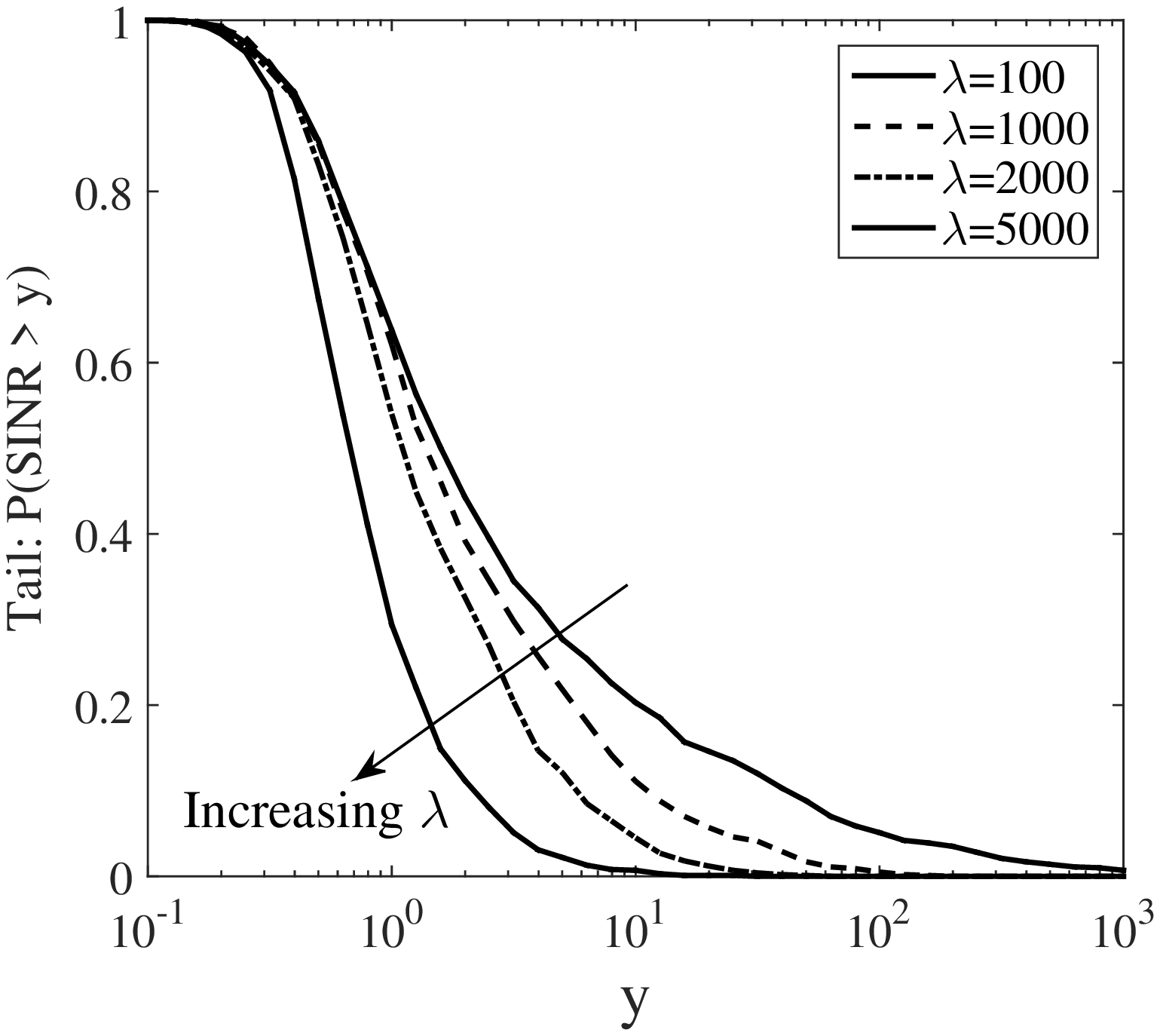}
		}
		\caption{Validation of the convergence of SINR to zero when $\tail_P \notin \mcalr_{-\alpha}$ with $\alpha \in [0,1]$. Simulated parameters include two-slope pathloss (i.e. $K =2$) with $A_0 = 1$, $\beta_1 = 4$, $R_1 = 10$ m, $R_{\infty} = 40$ km, two-dimensional network domain (i.e. $d=2$), and network density $\lambda$ in BSs/$\text{km}^2$.}
		\label{fig:ConvergYto0}
	\end{figure}
	
	In practically relevant network configurations, the pathloss attenuation is bounded (i.e. $\beta_0 = 0$) and channel power is normally less heavy-tailed or even truncated (i.e. $\tail_m = o(\bar{H})$ for some $\bar{H} \in \mcalr_{-\infty}$). In particular, the conventional case of lognormal shadowing and Rayleigh fading results in channel power belonging to the class $\mcalr_{-\infty}$. As a result, based on \thmref{thm:tailequiv}, we have that $\tail_P \in \mcalr_{-\infty}$, hence $\sinr \stackrel{a.s.}{\to} 0$. Therefore, the SINR is proven to be asymptotically decreasing with the infrastructure density. This means that there is a fundamental limit on network densification and the network should not operate in the ultra dense regime since deploying excessively many BSs would decrease the network performance due to the fact that signal power boosting cannot compensate for the faster growing aggregate interference (i.e. $M/I \stackrel{a.s.}{\to} 0$).  In other words, there exists an optimal density value until which the SINR monotonically increases, and after which the SINR monotonically decreases.
	
	In \fref{fig:ConvergYto0}, we provide simulation results with $\tail_P \notin \mcalr_{-\alpha}$ with $\alpha \in [0,1]$, with $\tail_P \in \mcalr_{-2}$ in \fref{fig:ConvergYto0}(a) and $\tail_P \in \mcalr_{-4}$ in \fref{fig:ConvergYto0}(b). We observe that the claim that the tail of SINR distribution vanishes and converges to zero when $\lambda$ increases is confirmed. The convergence of SINR to zero in the high density regime further emphasizes the cardinal importance of performing local scheduling among BSs, as well as signal processing mechanisms for interference mitigation.

\subsection{User Performance Scaling Regimes}
	
	We investigate now the fundamental limits to the amount of densification, which depend not only on the pathloss but also on the channel power distribution. Based on previous analytical results and the below theorem, we show that there exists a phase transition when the network density goes to infinity (ultra-densification). Depending on the pathloss attenuation (singularity and multi-slope) and the channel power distribution, there could be three distinct regimes for the coverage and the average rate: monotonically increasing, saturation, and deficit.
	
	\begin{thm}
		The coverage $\Pp_y(\lambda)$ for fixed $y$ and the rate $\Cc(\lambda)$ scale as follows:
		\begin{enumerate}
			\item $\Pp_y(\lambda) \to 1$ and $\Cc(\lambda) \to \infty$ as $\lambda \to \infty$ if $\tail_P \in \mcalr_0$;
			\item $\frac{\Pp_y(u\lambda)}{\Pp_y(\lambda)} \to 1$ and $\frac{\Cc(u\lambda)}{\Cc(\lambda)} \to 1$ for $0 < u < \infty$ as $\lambda \to \infty$ if $\tail_P \in \mcalr_{-\alpha}$ with $0 < \alpha < 1$;
			\item $\Pp_y(\lambda) \to 0$ and $\Cc(\lambda) \to 0$ as $\lambda \to \infty$ if $\tail_P \notin \mcalr_{-\alpha}$ with $\alpha \in [0,1]$; moreover there exist finite densities $\lambda_p, \lambda_c$ such that $\Pp_y(\lambda_p) > \lim_{\lambda \to \infty}\Pp_y(\lambda)$ and $\Cc(\lambda_c) > \lim_{\lambda \to \infty}\Cc(\lambda)$.
		\end{enumerate}
		\label{thm:PerfLimit}
	\end{thm}
	\begin{IEEEproof}
		\textnormal{See Appendix \ref{appen_Theorem3}.}
	\end{IEEEproof}
	
	The above result identifies three scaling regimes for the coverage probability and rate depending on the behavior of $\tail_P$:
	\begin{figure}[!t]
		\centering
		\subfigure[Coverage probability with $\beta_0=0$]
		{
			\includegraphics[width=0.475\textwidth]{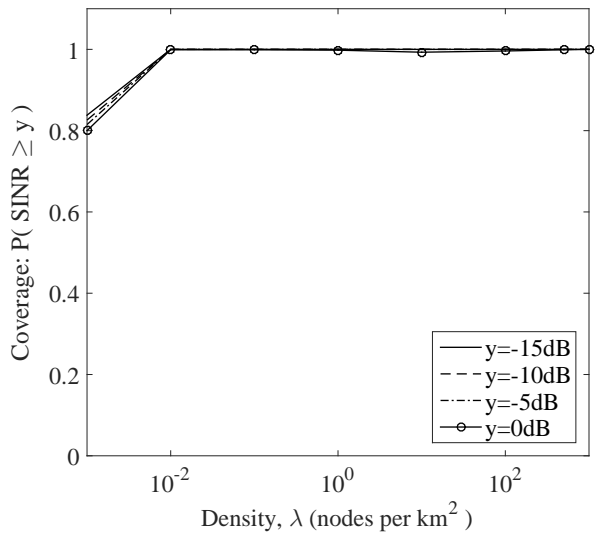}
		}
		\subfigure[Average rate with $\beta_0=0$]
		{
			\includegraphics[width=0.475\textwidth]{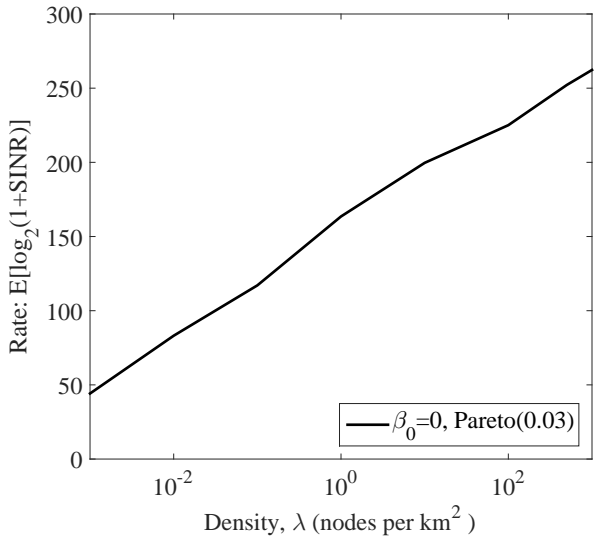}
		}
		\caption{Experiment of \emph{growth regime}. Approximation of $\tail_P \in \mcalr_0$ by $\tail_m \sim \Pareto(0.03)$ in Matlab simulation. Simulated parameters include two-slope pathloss (i.e. $K =2$) with $A_0 = 1$, $\beta_1 = 4$, $R_1 = 10$ m, $R_{\infty} = 40$ km, two-dimensional network domain (i.e. $d=2$), and network density $\lambda$ in BSs/$\text{km}^2$.}
		\label{fig:GrowRegime}
	\end{figure}
	\begin{itemize}
		\item \emph{Growth regime}: When $\tail_P \in \mcalr_0$, meaning that the channel power $m$ is slowly varying $\tail_m \in \mcalr_0$ (see \corref{cor:classifyTail}), both coverage and rate are monotonically increasing with $\lambda$. In particular, the average rate asymptotically grows with the network density. In \fref{fig:GrowRegime}, we show simulations  with $\tail_m \sim \Pareto(0.03)$\footnote{Pareto distribution with relatively small varying index $\alpha$ was used as an approximation of $\tail_m \in \mcalr_0$ since standard software packages (e.g. Matlab) do not have built-in tools for slowly varying distributions. Note that the smaller the varying index is, the more likely the realizations of the channel power are to have large values, resulting in numerical overflow.}. We can see that, even though the pathloss is bounded, the coverage probability is almost one for all SINR thresholds, while throughput increases almost linearly with the logarithm of the network density in the evaluated range. This growth regime, revealed from \thmref{thm:PerfLimit}, shows that the great expectations on the potential of network densification are theoretically possible. However, since slowly varying distributions are rather theoretical extremes and would be rarely observable in practice, the growth regime would be highly unlikely in real-world networks.
		
		\begin{figure}[!t]
			\centering
			\subfigure[Coverage probability with $\beta_0 = 3$]
			{
				\includegraphics[width=0.475\textwidth]{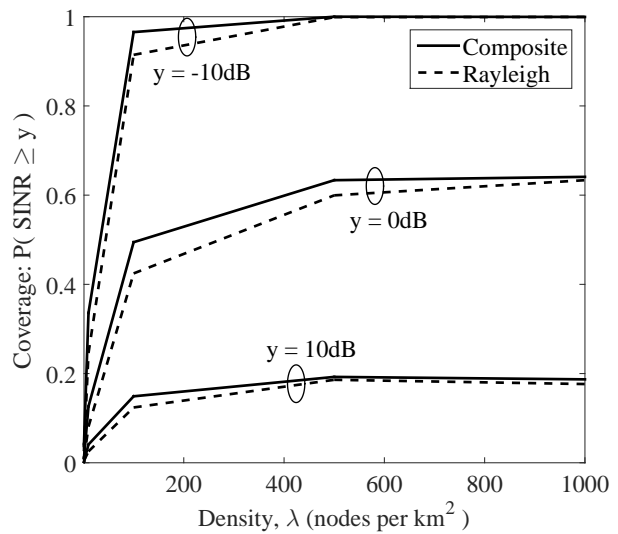}		
			}
			\subfigure[Average rate with $\beta_0 = 3$]
			{
				\includegraphics[width=0.47\textwidth]{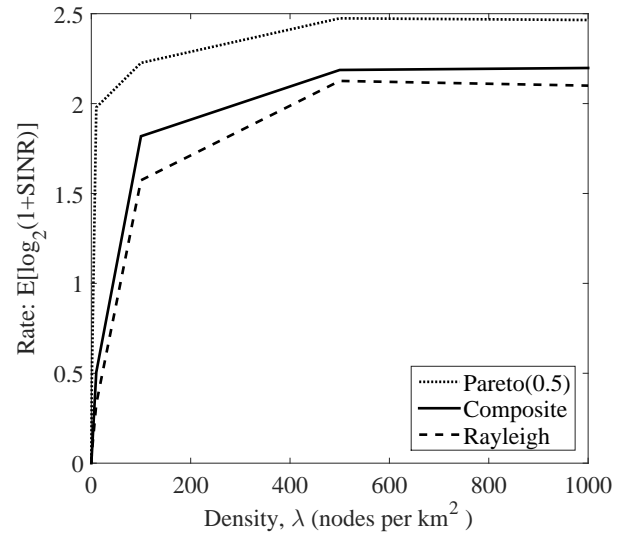}		
			}
			\caption{\emph{Saturation regime} due to $\beta_0 > d$. Here, \emph{saturation} of both coverage probability (left plot) and rate (right plot) is due to large near-field pathloss exponent $\beta_0 = 3$, which leads to $\tail_P \in \mcalr_{-\alpha}$ with $\alpha \in (0,1)$ for all considered types of the channel power distribution including composite Rayleigh-lognormal, Rayleigh, and Pareto(0.5). Simulated parameters include two-slope pathloss (i.e. $K =2$) with $A_0 = 1$, $\beta_1 = 4$, $R_1 = 10$ m, $R_{\infty} = 40$ km, two-dimensional network domain (i.e. $d=2$), and network density $\lambda$ in BSs/$\text{km}^2$.}
			\label{fig:SaturRegime}
		\end{figure}
		\begin{figure}[!t]
			\centering
			\subfigure[Coverage probability]
			{
				\includegraphics[width=0.475\textwidth]{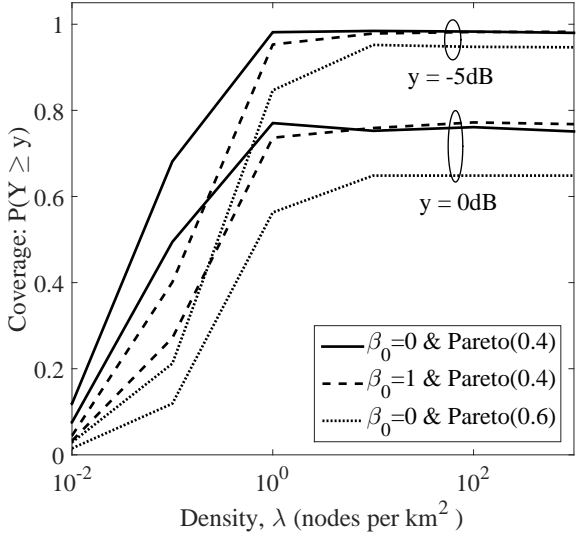}	
			}
			\subfigure[Average rate]
			{
				\includegraphics[width=0.475\textwidth]{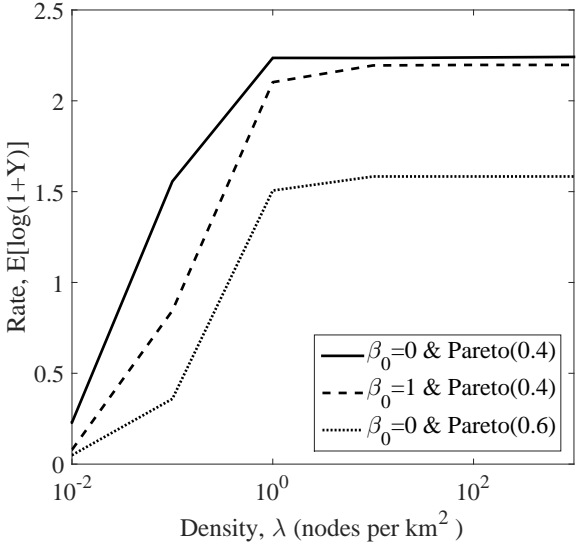}	
			}
			\caption{\emph{Saturation regime} due to \emph{regularly varying} channel power with index in $(-1,0)$. Here, \emph{saturation} of both coverage probability (left plot) and rate (right plot) is due to regularly varying channel power $\tail_m \sim \Pareto(\alpha)$ with $\alpha = 0.4$ and $\alpha = 0.6$. Simulated parameters include two-slope pathloss (i.e. $K =2$) with $A_0 = 1$, $\beta_1 = 4$, $R_1 = 10$ m, $R_{\infty} = 40$ km, two-dimensional network domain (i.e. $d=2$), and network density $\lambda$ in BSs/$\text{km}^2$.}
			\label{fig:SaturRegime2}
		\end{figure}
		\begin{figure}[!t]
			\centering
			\subfigure[Coverage probability, $\beta_0 < d$ and $\tail_m \sim \Composite$]
			{
				\includegraphics[width=0.475\textwidth]{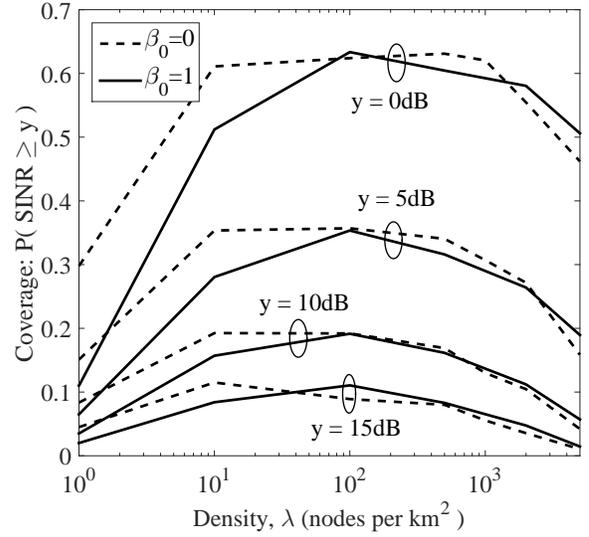}
			}
			\subfigure[Average rate, $\beta_0 < d$ and different $\tail_m$]
			{
				\includegraphics[width=0.475\textwidth]{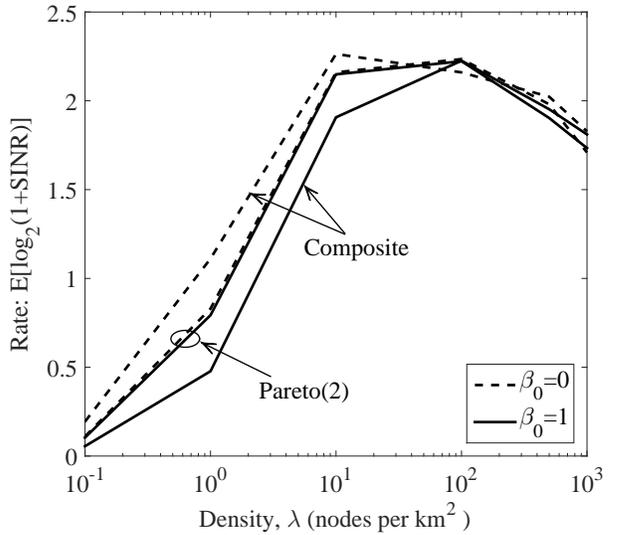}
			}
			\caption{Validation of \emph{deficit regime} when $\tail_P \notin \mcalr_{-\alpha}$ with $\alpha \in [0,1]$. Simulated parameters include two-slope pathloss (i.e. $K =2$) with $A_0 = 1$, $\beta_1 = 4$, $R_1 = 10$ m, $R_{\infty} = 40$ km, two-dimensional network domain (i.e. $d=2$), and network density $\lambda$ in BSs/$\text{km}^2$.}
			\label{fig:DeficitRegime}	
		\end{figure}		
		\item \emph{Saturation regime}: When the channel power is regularly varying with index within $-1$ and $0$ or the near-field pathloss exponent $\beta_0$ is larger than the network dimension $d$, the tail distribution of wireless link behaves as $\tail_P \in \mcalr_{-\alpha}$ with $\alpha \in (0,1)$. Consequently, both coverage probability and rate saturate past a certain network density. This saturation behavior is also confirmed by simulation experiments as shown in \fref{fig:SaturRegime} and \fref{fig:SaturRegime2}. In \fref{fig:SaturRegime}, it is the pathloss function's singularity with $\beta_0 > d$ that creates performance saturation for any type of channel power distribution not belonging to the class $\mcalr_0$. In \fref{fig:SaturRegime2}, the saturation is completely due to regularly varying channel power $\tail_m \in \mcalr_{-\alpha}$, $\alpha \in (0,1)$, regardless of pathloss boundedness. Prior studies have shown that the network performance is invariant of the network density for unbounded pathloss and negligible background noise. Our results (\thmref{thm:PerfLimit}) show that this performance saturation may happen in a much larger setting, including (i) with non-negligible thermal noise, and (ii) even with bounded pathloss if channel power is in the class $\mcalr_{-\alpha}$ with $\alpha \in (0,1)$. More importantly, the \emph{unbounded} property of the pathloss - widely used in the literature - is just a necessary condition to have saturated performance. A sufficient condition is that the near-field pathloss exponent has to be greater than the network dimension, i.e. $\beta_0 > d$. As we will see shortly, when channel power is less heavy-tailed than the class $\mcalr_{-\alpha}$ with $\alpha \in [0,1]$, then unbounded pathloss with $0 < \beta_0 < d$ results in the same scaling regime as bounded pathloss does.
		
		\item \emph{Deficit regime}: The third regime of network densification is determined first by a channel power distribution that is less heavy-tailed, precisely all remaining distributions not belonging to $\mcalr_{-\alpha}$ with $\alpha \in [0,1]$, and second by a near-field pathloss exponent smaller than the network dimension (i.e. $\beta_0 < d$). In this regime, both coverage probability and rate initially increase in the low density regime, then achieve a maximum at a finite network density, after which they start decaying and go to zero in the ultra dense regime. This behavior is also confirmed by simulations shown in \fref{fig:DeficitRegime}. More precisely, both coverage probability (left plot) and rate (right plot) exhibit `\emph{inverse U}' curves with respect to the network density $\lambda$ for different SINR thresholds $y$ and different types of channel power distribution. This suggests that there is an optimal point of network density to aim for. Particularly, this deficit regime can happen even with unbounded pathloss given that the near-field exponent is still smaller than the network dimension (i.e. $\beta_0 < d$); the bounded pathloss used in prior works ($\beta_0 = 0$) to obtain this deficit regime is a special case of this class.			
	\end{itemize}
	
	\tref{tab:PerfCases} summarizes the behavior of user performance according to \thmref{thm:PerfLimit} and shows the three different regimes. First, we see that the optimistic expectation of ever-growing user's rate and full coverage in UDNs is theoretically possible, though unlikely in reality. Second, we shed light on the divergence between previous results on the fundamental limits of network densification. It is due to two different assumptions on the pathloss model, one with $\beta_0 > d$ that results in saturation regime, and the other with $\beta_0=0$ that results in deficit regime.	

	% Table in two-column paper format
	{\renewcommand{\arraystretch}{1.5}
		\begin{table}[!t]
			\centering
			\caption{User Performance Scaling Regimes.}\label{tab:PerfCases}   
			\begin{tabular}{c||c|c|c|c}
				\hline\hline
				\multirow{2}{*}{\parbox[m]{1.2cm}{Scaling regime}} & \multicolumn{3}{c|}{$\tail_m \in \mcalr_{-\alpha}$} & \multirow{2}{*}{\parbox[m]{0.9cm}{\centering lighter tail}} \\
				\cline{2-4}
				& \parbox[m]{0.9cm}{\centering $\alpha = 0$} & \parbox[m]{1.6cm}{\centering $0 < \alpha < 1$} & \parbox[m]{0.9cm}{\centering $\alpha > 1$} &  \\
				\hline\hline
				$\beta_0 < d$ & \multirow{2}{*}{\parbox[m]{1.9cm}{\centering $\sinr \overset{p}{\to} \infty$, \\ $\Pp \to 1$, \\ $\Cc \to \infty$}} & \multirow{2}{*}{saturation} & \multicolumn{2}{c}{inverse U}  \\
				\cline{1-1}\cline{4-5}
				$\beta_0 > d$ &  &  & \multicolumn{2}{c}{saturation}\\
				\hline
			\end{tabular}
		\end{table}
	}
			
\subsection{Network Performance Scaling Regimes}

	Using \thmref{thm:PerfLimit}, we can now obtain the scaling laws of network-level performance metrics.
	
	\begin{cor}
		The coverage density $\Dc_y$ for fixed $y > 0$, and the area spectral efficiency $\Ac$ scale as follows:
		\begin{enumerate}
			\item $\frac{\Dc_y(\lambda)}{\lambda} \to 1$ and $\frac{\Ac(\lambda)}{\lambda} \to \infty$ as $\lambda \to \infty$ if $\tail_P \in \mcalr_0$;
			\item $\frac{\Dc_y(\lambda)}{\lambda} \to c_{y,\alpha}$ and $\frac{\Ac(\lambda)}{\lambda} \to c_{\alpha}$ as $\lambda \to \infty$ if $\tail_P \in \mcalr_{-\alpha}$ with $0 < \alpha < 1$, where constant $c_{y,\alpha} \in [0,1]$ depending on $y$ and $\alpha$, and constant $c_{\alpha} > 0$ depending on $\alpha$;
			\item $\frac{\Dc_y(\lambda)}{\lambda} \to 0$ and $\frac{\Ac(\lambda)}{\lambda} \to 0$ as $\lambda \to \infty$ if $\tail_P \notin \mcalr_{-\alpha}$ with $\alpha \in [0,1]$.
		\end{enumerate}
		\label{cor:NwkLimit}
	\end{cor}
	\begin{IEEEproof} See \aref{appen_corNwkLimit}.	\end{IEEEproof}
	
	We can easily see from \corref{cor:NwkLimit} that the network-level performance scales with the network infrastructure more optimistically than the user performance does. In particular, both the \emph{growth} and \emph{saturation} regimes of user performance result in \emph{growth} regime of network performance. 
	
	In the \emph{deficit} regime of user performance, the network performance scales sub-linearly with the network density. Nevertheless, this does not necessarily mean that $\Dc_y(\lambda)$ and $\Ac(\lambda)$ always vanish in the same manner as $\Pp_y$ and $\Cc$ do when $\lambda \to \infty$. It depends on the precise tail behavior of $\tail_P$; and in the case that the resulting $\Pb(\sinr > y)$ does not vanish faster than $\lambda^{-\varepsilon}$ for some $0 < \varepsilon < 1$ as $\lambda \to \infty$, then $\Dc_y$ and $\Ac$ will be in the same order of $\lambda^{1-\varepsilon}$ as $\lambda \to \infty$. In such cases, $\Dc_y$ and $\Ac$ may still increase with the network infrastructure although they will increase with a much lower speed, i.e. $\Dc_y(\lambda) = o(\lambda)$ and $\Ac(\lambda) = o(\lambda)$ as $\lambda \to \infty$.

	\subsection{Ordering Results and Design Guidelines}
	The mathematical framework developed above characterizes the asymptotic behavior of coverage and throughput in dense spatial networks. In short, knowing the tail behavior of fading and pathloss characteristics, we can see in which regime (growth, saturation or deficit) the performance falls in. We want now to compare two networks with same scaling regime but different characteristics (e.g. different number of transmit antennas, different shadowing, etc.) and see where we will achieve better network performance. One approach to answer this question would be to derive the performance metrics in closed form. Although the exact distribution of $\sinr$ is known for a simplified network model, e.g. \cite{Nguyen2010,Andrews2011}, it is still analytically cumbersome for realistic and practically relevant system models. In the absence of handy analytical expressions, it is difficult to compare different transmission techniques in general network settings. For that, in additional to the scaling laws, we develop ordering results for both coverage and average rate in order to facilitate more fine comparison between UDNs with different parameters. The ordering approach provides crisp insights and useful design guidelines into the relative performance of different transmission techniques, while circumventing the need to evaluate complicated coverage and rate expressions. As such, the scaling results provide an answer on the asymptotic performance behavior (increase, saturation, or inverse-U), while the ordering results aim at identifying when we have  superior performance in terms of coverage and rate among networks with the same asymptotic performance.
	
	We derive first the asymptotic Laplace transform of the inverse of SINR, denoted by $Z = 1/\sinr$. 
	\begin{lem}\label{lem:LaplaceZ}
		If $\tail_P \in \mcalr_{-\alpha}$ with $\alpha \in [0,\infty)$, then for $s \in \R^+$,
		\begin{equation*}
			\Lc_{Z}(s) = \left(1+ \alpha\int_0^1 (1 - e^{-st})\frac{\d{t}}{t^{\alpha+1}} \right)^{-1}, \quad \text{ as } \lambda \to \infty.
		\end{equation*}
	\end{lem}
	\begin{IEEEproof}
		\textnormal{See Appendix \ref{appen_Lemma4}.}
	\end{IEEEproof}	
	
	\begin{thm}\label{thm:Ordering}
		For two networks $\Xi_1$ and $\Xi_2$ with the same density $\lambda$ and with distribution of wireless link $F_1$ and $F_2$, resp., if $\tail_1 \in \mcalr_{-\alpha_1}$, $\tail_2 \in \mcalr_{-\alpha_2}$, and $0 \leq \alpha_1 \leq \alpha_2 \leq \infty$, then
		\begin{align*}
			\E(\sinr_1) & \geq \E(\sinr_2), \\
			\E(\log(1+\sinr_1)) & \geq \E(\log(1+\sinr_2)),
		\end{align*}
		as $\lambda \to \infty$, where $\sinr_1$ and $\sinr_2$ are the received SINR of $\Xi_1$ and $\Xi_2$, respectively.
	\end{thm}
	\begin{IEEEproof}
		\textnormal{See Appendix \ref{appen_Theorem4}.}
	\end{IEEEproof}
	
	\thmref{thm:Ordering} states that the heavier the tail of wireless link distribution, the better the performance under ultra-densification. As shown by \thmref{thm:tailequiv}, a heavier-tailed wireless link distribution can be obtained through either a heavier-tailed channel power distribution $m$ or greater near-field pathloss exponent $\beta_0$. Physically, heavier-tailed channel power $m$ means greater probability of high channel power, taking into account all effects, such as transmit power, array gains, beamforming gain. Since higher channel power may also increase interference, a natural question is whether it is beneficial to have higher beamforming gain (or higher channel power in general) in the high density regime. Note that large beamforming gains will increase the interference towards the users who are located in the beam-direction of the intended user, however since large beamforming gains are achieved with narrow beams, this probability may be low. Here, \thmref{thm:Ordering} states that that achieving higher beamforming gains or using techniques that render the tail of wireless link distribution heavier are beneficial in terms of network performance. In short, directional transmissions can be beneficial for the network performance, as are envisioned for massive MIMO and millimeter wave systems.

%%%%%%%%%%%%%%%%%%%%%%%%%%%%%%%%%%%%%%%%%%%%%%%%%%%%%%%%%%%%%%%%%%
\section{Conclusions}\label{s:Conclusion}
	This paper has presented a new framework for analyzing the performance - in terms of downlink SINR, coverage, and throughput - of wireless network densification and for identifying its potential and challenges. We considered a practically relevant channel model that captures multi-slope pathloss and general channel power distributions, including transmit power, shadowing, fast fading, as well as associated gains such as antenna pattern and beamforming gain. A key finding is that under strongest cell association, the performance of ultra dense wireless networks exhibits three distinct regimes of the user performance, namely growth, saturation, and deficit regime. The tail behavior of the channel power and near-field pathloss exponent are the key parameters that determine the performance limits and the asymptotic scaling. Some particular implications include:
	\begin{itemize}
		\item Monotonically increasing per-user performance (coverage probability and average rate) by means of ultra-densification is theoretically possible, though highly unlike in reality since it requires slowly varying tail of the channel power distribution, which is a theoretical extreme.
		\item In practice, installing more BSs is beneficial to the user performance up to a density point, after which further densification can become harmful user performance due to faster growth of interference compared to useful signal. This highlights the cardinal importance of interference mitigation, coordination among neighboring cells and local spatial scheduling.
		\item The network performance in terms of coverage density and area spectral efficiency benefit from the network densification more than user performance does. In particular, it scales linearly with the network infrastructure when the user performance is in growth or saturation regime.
		\item Increasing the tail distribution of the channel power using advanced transmission techniques, such as massive MIMO, CoMP, and directional beamforming, is beneficial as it improves the performance scaling regime. Moreover, the effect of emerging technologies (e.g. D2D, mmWave) on near-field pathloss and channel power distribution need to be studied.
		\item It is meaningful to determine the optimal network density beyond which further densification becomes destructive or cost-ineffective. This operating point will depend on properties of the channel power distribution, noise level, and pathloss in the near-field region, and is of cardinal importance for the successful deployment of 5G UDNs. 
	\end{itemize}

%%%%%%%%%%%%%%%%%%%%%%%%%%%%%%%%%%%%%%%%%%%%%%%%%%%%%%%%%%%%%%%%%%
\appendix
\appendices

\subsection{Distribution of Wireless Link}	
	We first investigate the exact distribution of the wireless link in a given wireless network $\Xi$.
	\begin{lem}\label{lem:DistanceCDF}
		If  $R_{\infty} < \infty$, then the distance from the user to a random node admits a non-degenerate distribution given as $G(r) = \pren{r/R_{\infty}}^d$ for $r \in [0,R_{\infty}]$, and $G(r) = 1$ for $r \geq R_{\infty}$.
	\end{lem}
	\begin{IEEEproof}
		Under the assumption that nodes are distributed according to a homogeneous PPP, the nodes of a realization $\phi$ of $\Phi$ are uniformly distributed. % in the network domain. 
		Thus, given $\phi$ and the assumption that $R_{\infty} < \infty$, the distribution of the distance to a random node of $\phi$, say $G(\cdot; \phi)$, is $G(r;\phi) = \pren{r/R_{\infty}}^d$ for $0 \leq r \leq R_{\infty}$, and $G(r;\phi) = 1$ for $r \geq R_{\infty}$. Then, taking $G(r) = \E_{\phi}(G(r;\phi))$, the result follows.
	\end{IEEEproof}
	\begin{remark*}
		First, note that the distribution $G$ is different from the usual \emph{void probability}, which is the distance to the closest node (nearest neighbor). Second, we can see that for unbounded network domains ($R_{\infty} = \infty$), the distance to a random node does not have a non-degenerate distribution. This is because under the PPP assumption, the number of nodes at equal distance increases with the circumference, which tends to infinity when the outer distance tends to infinity, leading to an absorption of probability. Therefore, using a limited network domain is not only more realistic, but also useful to have a normally behaving distribution of the distance.
	\end{remark*}
	
	\begin{lem}\label{lem:tailP}
		Denote $a_{K} = A_{K-1} R_{\infty}^{\beta_{K-1}}$, and for $k = 0,\ldots, K-1$ denote $a_k = A_k R_k^{\beta_k}$, and
		\begin{equation}\label{eq:Jk}
		J_k(t) = \frac{\E\pren{m^{\delta_k} \one\pren{a_k t \leq m < a_{k+1} t}}}{A_k^{\delta_k} R_{\infty}^d} t^{-\delta_k}.
		\end{equation}
		Then,
		\begin{equation}\label{eq:tailP}
		\tail_P(t) = \tail_m\pren{a_K t} + \sum_{k=k_0}^{K-1}J_k(t),
		\end{equation}
		where $k_0=0$ for $\beta_0 > 0$, and $k_0=1$ for $\beta_0 = 0$.
	\end{lem}

	\begin{IEEEproof}
		% We have
		%    \begin{equation*}%\label{eq:tailP_base}
		%        \tail_P(t) = \Pb(P \geq t) = \Pb\pren{\frac{m}{l(r)} \geq t} = \E_r\pren{\tail_m(t l(r))}.
		%    \end{equation*}
		Using $l(r)$ given in \eref{eq:PL}, we have
		\begin{equation}\label{eq:tailP_expr}
		\tail_P(t) = \int_{0}^{R_{\infty}}\tail_m(tl(r))G(\d{r}) = \sum_{k=0}^{K-1} \mathcal{I}_k(t),
		\end{equation}
		where
		%\begin{equation*}
		% \mathcal{I}_k(t) := \int_{R_k}^{R_{k+1}}\tail_m\pren{A_k r^{\beta_k} t} G(\d{r}).
		%\end{equation*}
		$\mathcal{I}_k(t) := \int_{R_k}^{R_{k+1}}\tail_m\pren{A_k r^{\beta_k} t} G(\d{r})$.
		
		(i) For $\beta_k > 0$: integration by parts with $\tail_m(A_k r^{\beta_k} t)$ and $G(\d{r})$ for $\mathcal{I}_k$ yields
		\begin{equation*}
		\mathcal{I}_k(t) = \tail_m\pren{A_k r^{\beta_k} t} G(r) \Big|_{R_k}^{R_{k+1}} + J_k(t),
		\end{equation*}
		where
		% \begin{equation*}
		%    J_k(t) = \int_{R_k}^{R_{k+1}}G(r)\d{F_m(A_k r^{\beta_k} t}),
		% \end{equation*}
		$J_k(t) = \int_{R_k}^{R_{k+1}}G(r)\d{F_m(A_k r^{\beta_k} t})$, 
		which reduces to \eref{eq:Jk} after change of variable $u = A_k r^{\beta_k} t$ and applying the condition $A_k R_{k+1}^{\beta_k} = A_{k+1} R_{k+1}^{\beta_{k+1}} = a_{k+1}$ due to \eref{eq:plScale}. On the other hand, applying again \eref{eq:plScale},
		\begin{equation}\label{eq:tailP_sum1}
		\sum_{k=0}^{K-1}\pren{\tail_m\pren{A_k r^{\beta_k} t} G(r) \Big|_{R_k}^{R_{k+1}}} %= \tail_m\pren{A_{K-1} R_{\infty}^{\beta_{K-1}} t}G(R_{\infty}) - \tail_m\pren{A_0 R_0^{\beta_0} t} G(0) 
		= \tail_m(a_K t).
		\end{equation}
		Hence, substituting \eref{eq:tailP_sum1} back to \eref{eq:tailP_expr} yields \eref{eq:tailP} with $k_0=0$.
		
		(ii) For $\beta_0 = 0$, we have
		\begin{equation*}
		\tail_P(t) = \mathcal{I}_0(t) + \sum_{k=1}^{K-1}\mathcal{I}_k(t) = 
		\int_{0}^{R_1}\tail_m\pren{A_0 t} G(\d{r}) + \sum_{k=1}^{K-1}\mathcal{I}_k(t).
		\end{equation*}
		Thus,
		%	\begin{equation*}
		%		\tail_P(t) = \tail_m(A_0 t) G(R_1) + \tail_m\pren{A_{K-1} R_{\infty}^{\beta_{K-1}} t}G(R_{\infty}) 		- \tail_m\pren{A_{1} R_{1}^{\beta_1} t}G(R_1) + \sum_{k=1}^{K-1}J_k(t),
		%	\end{equation*}
		\begin{multline*}
		\tail_P(t) = \tail_m(A_0 t) G(R_1) + \tail_m\pren{A_{K-1} R_{\infty}^{\beta_{K-1}} t}G(R_{\infty}) \\
		- \tail_m\pren{A_{1} R_{1}^{\beta_1} t}G(R_1) + \sum_{k=1}^{K-1}J_k(t),
		\end{multline*}
		where $A_0 = A_{1} R_{1}^{\beta_1}$ due to \eref{eq:plScale}. Hence, \eref{eq:tailP} with $k_0=1$.	
	\end{IEEEproof}

\subsection{Proof of \thmref{thm:tailequiv}}\label{appen_Theorem1}
	(1a) For the case of $\tail_m \in \mcalr_{-\alpha}$ for $\alpha \in [0,\infty)$, $\tail_m$ can be represented as $\tail_m(x) \sim x^{-\alpha} L(x)$ for some $L \in \mcalr_0$. Then, by the monotone density theorem (see \lemref{lem:MonotoneDensity} and \cite{Bingham1989,Embrechts1997}), the density function $f_m$ of $F_m$ can be written as
	% \begin{equation*}
	%    f_m(t) \sim \alpha t^{-\alpha-1}L(t), \quad \text{ as } t \to \infty.
	% \end{equation*}
	$f_m(t) \sim \alpha t^{-\alpha-1}L(t)$ as $t \to \infty$.
	
	\underline{\emph{For $k \geq 1$}}: As $t \to \infty$, \[\E\pren{m^{\delta_k}\one(a_k t \leq m < a_{k+1} t)} \sim \int_{a_k t}^{a_{k+1}t}\alpha x^{\delta_k - \alpha - 1} L(x) \d{x}.\]
	
	\noindent If $\alpha < \delta_k$: by Karamata's theorem (see \lemref{lem:KaramataThm} and \cite[Prop.~1.5.8]{Bingham1989}), for $t_0 > 0$:
	\begin{align*}
	& \int_{a_k t}^{a_{k+1}t}\alpha x^{\delta_k - \alpha - 1} L(x) \d{x} \\
	&= \int_{t_0}^{a_{k+1}t}\alpha x^{\delta_k - \alpha - 1} L(x) \d{x} - \int_{t_0}^{a_{k}t}\alpha x^{\delta_k - \alpha - 1} L(x) \d{x} \nonumber \\
	&\sim \frac{\alpha}{\delta_k - \alpha}\pren{a_{k+1}^{\delta_k - \alpha} - a_k^{\delta_k - \alpha}} L(t) t^{\delta_k - \alpha}, \quad \text{ as } t \to \infty.
	\end{align*}
	If $\alpha > \delta_k$: similarly to the above case, we can easily obtain the same result using Karamata's theorem. Note that if $\alpha = \delta_{\hat{k}}$ for some $\hat{k} \in [1,K-1]$, we can also easily show that $J_{\hat{k}}(t) \sim c L(t) t^{-\alpha}$ for some constant $c$. Thus, for $k \geq 1$,
	\begin{align}
	J_k(t) %& = \frac{\E\pren{m^{\delta_k} \one(a_k t \leq m < a_{k+1}t)}}{A_k^{\delta_k} R_{\infty}^d t^{\delta_k}} \nonumber \\
	\sim \frac{\alpha}{\delta_k - \alpha} \frac{a_{k+1}^{\delta_k - \alpha} - a_k^{\delta_k - \alpha}}{A_k^{\delta_k}R_{\infty}^d} L(t) t^{-\alpha}, \text{ as } t \to \infty. \label{eq:proofTailEquivk1}
	\end{align}
	
	\underline{\emph{For $k=0$}}: If $\delta_0 < \alpha$, then $\E(m^{\delta_0})$ exists and $\E\pren{m^{\delta_0}\one(0 \leq m < a_{1} t)} = \E(m^{\delta_0})$ as $t \to \infty$. Thus, $J_0(t) = \E(m^{\delta_0})A_0^{-\delta_0} R_{\infty}^{-d} t^{-\delta_0}$ as $t \to \infty$. Otherwise, i.e. if $\delta_0 \geq \alpha$, then we have
	%	\begin{equation*}
	%		\E\pren{m^{\delta_0}\one(0 \leq m < a_{1} t)} = \int_{0}^{a_1t}x^{\delta_0}F_m(\d{x}) \overset{(a)}{=} \frac{\alpha (a_1 t)^{\delta_0}}{\delta_0 - \alpha} \tail_m(a_1t) \overset{(b)}{=} \frac{\alpha (a_1 t)^{\delta_0 - \alpha}}{\delta_0 - \alpha} L(t),
	%	\end{equation*}
	\begin{multline*}
	\E\pren{m^{\delta_0}\one(0 \leq m < a_{1} t)} = \int_{0}^{a_1t}x^{\delta_0}F_m(\d{x}) = \\
	\overset{(a)}{=} \frac{\alpha (a_1 t)^{\delta_0}}{\delta_0 - \alpha} \tail_m(a_1t) \overset{(b)}{=} \frac{\alpha (a_1 t)^{\delta_0 - \alpha}}{\delta_0 - \alpha} L(t),
	\end{multline*}
	where $(a)$ is due to \cite[Prop.~A3.8]{Embrechts1997}, and $(b)$ is due to the representation $\tail_m(x) \sim x^{-\alpha} L(x)$. Thus, as $t \to \infty$,
	\begin{align}
	J_0(t) & = \frac{\E_m\pren{m^{\delta_0}\one(0 \leq m < a_{1} t)}}{A_0^{\delta_0}R_{\infty}^d t^{\delta_0}} \nonumber\\
	& = \begin{cases} \E(m^{\delta_0}) A_0^{-\delta_0}R_{\infty}^{-d} t^{-\delta_0}, & \text{ if } \delta_0 < \alpha \\ \alpha (\delta_0 - \alpha)^{-1} a_1^{\delta_0-\alpha} L(t) t^{-\alpha}, & \text{ if } \delta_0 \geq \alpha \end{cases}.\label{eq:proofTailEquivk0}
	\end{align}    
	By substituting \eref{eq:proofTailEquivk1} and \eref{eq:proofTailEquivk0} into the expressions of $\tail_P$ given by \lemref{lem:tailP}, we have, as $t \to \infty$,
	\begin{align*}
	& \tail_P(t) = \tail_m\pren{a_K t} + \sum_{k=i}^{K-1}J_k(t) \\
	& \sim \begin{cases} \pren{a_K^{-\alpha} + \sum_{k=1}^{K-1} C_k } \frac{L(t)}{t^{\alpha}}, \quad \text{if } \beta_0 = 0,\\
	\pren{a_K^{-\alpha} + \sum_{k=0}^{K-1} C_k} \frac{L(t)}{t^{\alpha}}, \quad \text{if } \beta_0 > 0, \delta_0 \geq \alpha \\
	\pren{a_K^{-\alpha} + \sum_{k=1}^{K-1} C_k} \frac{L(t)}{t^{\alpha}} + \frac{\E(m^{\delta_0})}{A_0^{\delta_0} R_{\infty}^d t^{\delta_0}}, \text{if } \beta_0 > 0, \delta_0 < \alpha,
	\end{cases}
	\end{align*}
	where $C_k = \alpha(a_{k+1}^{\delta_k - \alpha} - a_k^{\delta_k - \alpha})\pren{(\delta_k - \alpha)A_k^{\delta_k}R_{\infty}^d}^{-1}$,
	and where for the last case with $\beta_0 > 0$ and $\delta_0 < \alpha$, we further have $t^{-\alpha} = o(t^{-\delta_0})$ as $t \to \infty$. As a result, $\tail_P$ is regularly varying with index $\alpha$ if $\alpha \leq \delta_0$, and with index $\delta_0$ if $\delta_0 < \alpha$. Hence the proof for $\tail_m \in \mcalr_{-\alpha}$ with $\alpha \in [0,\infty)$.
	
	(1b) Now, assume that $F_m \in \mcalr_{-\infty}$.
	For $k \geq 1$, we have
	\begin{align}
	\E&\pren{m^{\delta_k} \one\pren{a_k t \leq m < a_{k+1}t}} 
	= \int_{a_k t}^{a_{k+1}t}x^{\delta_k}\d{F_m(x)} \nonumber\\
	& = \frac{\tail_m(a_k t)}{a_k^{-\delta_k} t^{-\delta_k}} - \frac{\tail_m(a_{k+1} t)}{a_{k+1}^{-\delta_k} t^{-\delta_k}} + \int_{a_k t}^{a_{k+1}t} \frac{\delta_k\tail_m(x)}{x^{1-\delta_k}}\d{x} \nonumber\\ %\brac{\tail_m(x) x^{\delta_k}}\Big|_{a_{k+1}t}^{a_{k}t}
	&\overset{(\ast)}{\sim} a_k^{\delta_k}\tail_m(a_k t) t^{\delta_k}, \quad \text{ as } t \to \infty, \label{eq:tailP_proof_case2_1}
	\end{align}
	where $(\ast)$ is by the facts that, as $t \to \infty$:
	\begin{itemize}
		\item $\tail_m(a_{k+1}t) = o(\tail_m(a_{k}t))$ since $\tail_m \in \mcalr_{-\infty}$ and $a_{k+1} > a_k$,
		\item and $\int_t^{\infty}x^{u}\tail_m(x)\d{x} = o(t^{u+1}\tail_m(t))$, $\forall{u}$, since $\tail_m \in \mcalr_{-\infty}$, see \lemref{lem:RapidVary}.% \cite[Theorem~A3.12]{Embrechts1997}.
	\end{itemize}
	Thus, for $k \geq 1$, as $t \to \infty$,
	\begin{equation}\label{eq:tailP_proof_case2_2}
	J_k(t) \sim \frac{a_k^{\delta_k}}{A_k^{\delta_k}R_{\infty}^d} \tail_m(a_k t) = \pren{\frac{R_k}{R_{\infty}}}^d \tail_m(a_k t).
	\end{equation}
	In addition, for $\beta_0 > 0$, as $t \to \infty$,
	\begin{equation}\label{eq:tailP_proof_case2_3}
	J_0(t) = \frac{\E_m\pren{m^{\delta_0} \one\pren{0 \leq m < a_1 t}}}{A_0^{\delta_0}R_{\infty}^d t^{\delta_0}} \to \frac{\E\pren{m^{\delta_0}}}{A_0^{\delta_0}R_{\infty}^d t^{\delta_0}}.
	\end{equation}
	Therefore,
	\begin{itemize}
		\item For $\beta_0 > 0$: 
		\begin{align*}
		\tail_P(t) & \sim \sum_{k=1}^{K}\pren{\frac{R_k}{R_{\infty}}}^d \tail_m(a_k t) + \frac{\E\pren{m^{\delta_0}}}{A_0^{\delta_0}R_{\infty}^d}t^{-\delta_0} \\
		& \sim \frac{\E\pren{m^{\delta_0}}}{A_0^{\delta_0}R_{\infty}^d}t^{-\delta_0}, \textrm{ as } t \to \infty. 
		\end{align*}
		Thus, $\tail_P \in \mcalr_{-\delta_0}$.
		\item For $\beta_0 = 0$:
		\begin{align*}
		\lim_{t \to \infty}\frac{\tail_P(ut)}{\tail_P(t)}
		& = \lim_{t \to \infty}\frac{\sum_{k=1}^{K}\pren{\frac{R_k}{R_{\infty}}}^d \tail_m(a_k u t)}{\sum_{k=1}^{K}\pren{\frac{R_k}{R_{\infty}}}^d \tail_m(a_k t)} \\
		& = \begin{cases} 0, & \text{ if } u > 1 \\
		\infty, & \text{ if } 0 < u < 1 \end{cases}.
		\end{align*}  
		Thus, $\tail_P \in \mcalr_{-\infty}$. In addition, since $\tail_m \in \mcalr_{-\infty}$ and $a_k < a_{k+1}$, we have $\tail_P(t) \sim (R_1/R_{\infty})^d \tail_m(A_0 t)$ where $A_0$ is due to the fact that $a_1 = A_1 R_1^{\beta_1} = A_0 R_1^{\beta_0} = A_0$ for $\beta_0 = 0$. This completes the proof of the first assertion.		
	\end{itemize}
	
	(2) Assume that $\tail_m(t) = o(\bar{H}(t))$ as $t \to \infty$ for $\bar{H} \in \mcalr_{-\infty}$. This means that $\tail_m$ decays more rapidly than $\bar{H}$, which in its turn decays the fastest among the class of the regularly varying distributions. Thus, \eref{eq:tailP_proof_case2_1} applies, and so \eref{eq:tailP_proof_case2_2} does. In addition, for $\beta_0 > 0$, we also have \eref{eq:tailP_proof_case2_3}. Thus, as $t \to \infty$,
	\begin{itemize}
		\item For $\beta_0 > 0$:
		%	\begin{equation*}
		$\tail_P(t) \sim \sum_{k=1}^{K}\pren{\frac{R_k}{R_{\infty}}}^d \tail_m(a_k t) + \frac{\E\pren{m^{\delta_0}}}{A_0^{\delta_0}R_{\infty}^d}t^{-\delta_0}
		\sim \frac{\E\pren{m^{\delta_0}}}{A_0^{\delta_0}R_{\infty}^d}t^{-\delta_0}.$
		%	\end{equation*}
		Thus, $\tail_P \in \mcalr_{-\delta_0}$.
		\item For $\beta_0 = 0$:
		%	\begin{equation*}
		$\tail_P(t) \sim \sum_{k=1}^{K-1}\pren{\frac{R_k}{R_{\infty}}}^d \tail_m(a_kt) \sim \pren{\frac{R_1}{R_{\infty}}}^d \tail_m(a_1 t)
		\overset{(\ast)}{=} \pren{\frac{R_1}{R_{\infty}}}^d \tail_m(A_0 t),$
		%	\end{equation*}
		where $(\ast)$ is by the fact that $a_1 = A_1 R_1^{\beta_1} = A_0 R_1^{\beta_0} = A_0$. Thus, $\tail_P(t) \sim \tail_m(A_0 t)$.
	\end{itemize}
	This proves the second assertion.

\subsection{Proof of \thmref{thm:Yinfty}}\label{appen_Theorem2}
	For $y \geq 0$, we have $\Pb(\sinr > y) = \Pb\pren{\frac{I - M + W}{M} < \frac{1}{y}} \to \Pb\pren{\frac{I}{M} - 1 < \frac{1}{y}}$ as $\lambda \to \infty$ since $W$ is finite and $m_i$ is not identical to zero.
	
	If $\tail_P \in \mcalr_0$, $I/M \overset{p}{\to} 1$ due to \cite{Maller1984}. Thus, $\forall y$
	\[\Pb(\sinr > y) = \Pb\pren{(I/M) - 1 < y^{-1}} = 1, \quad \text{as }\lambda \to \infty.\]
	
	If $\tail_P \in \mcalr_{-\alpha}$ with $0 < \alpha < 1$, $M/I \overset{d}{\to} R$ as $\lambda \to \infty$ where $R$ has a non-degenerate distribution \cite{Bingham1981}. As a result,
	\[\Pb(\sinr > y) = \Pb\pren{(I/M) < 1 + y^{-1}} \to D, \quad \text{as }\lambda \to \infty,\]
	where $D$ is a non-degenerate distribution.
	
	If $\tail_P \in \mcalr_{-\alpha}$ with $\alpha > 1$ or $\tail_P = o(\bar{H})$ with $\bar{H} \in \mcalr_{-\infty}$, we have $\E(P) < \infty$. Hence, $M/I \overset{a.s.}{\to} 0$ due to \cite{OBrien1980}. Moreover, $\forall y \in (0,\infty)$
	\begin{equation*}
	\Pb\pren{\sinr > y} = \Pb\pren{\frac{M}{I+W} > \frac{y}{1+y}} \leq \Pb\pren{\frac{M}{I} > \frac{y}{1+y}}.
	\end{equation*}
	Thus, $M/I \overset{a.s.}{\to} 0$ leads to $\sinr \overset{a.s.}{\to} 0$ as $\lambda \to \infty$.

\subsection{Proof of \thmref{thm:PerfLimit}}\label{appen_Theorem3}
	We start by showing that in sparse networks, the signal quality is improved when the node density increases.
	\begin{lem}\label{lem:Yzero}
		Let $0 \leq \lambda_1 < \lambda_2$. If $W > 0$, then $\sinr(\lambda_2) \overset{st}{>} \sinr(\lambda_1)$ as $\lambda_2 \to 0^+$.
	\end{lem}
	\begin{IEEEproof}
		As $\lambda \to 0^+$, we have $(I(\lambda)-M(\lambda)) = o(W)$ almost surely. Thus, for $y > 0$,
		\begin{align*}	    
		\lim_{\lambda_2 \downarrow 0}\Pb\pren{\sinr(\lambda_2) \geq y} & = \lim_{\lambda_2 \downarrow 0}\Pb\pren{M(\lambda_2) \geq y W} \\
		& \overset{(a)}{>} \lim_{\lambda_1 < \lambda_2 \downarrow 0}\Pb\pren{M(\lambda_1) \geq y W} \\
		& = \lim_{\lambda_1 < \lambda_2 \downarrow 0}\Pb\pren{\sinr(\lambda_1) \geq y},
		\end{align*} 
		where note that $(a)$ is intuitively evident, but a formal proof can be easily obtained using for example \cite[Prop.~2.4.2]{Baccelli2009}.
	\end{IEEEproof}
	
	Proof of \thmref{thm:PerfLimit} using \lemref{lem:Yzero} is as follows. Case (1) directly follows \thmref{thm:Yinfty}. % and $\sinr \overset{p}{\to} \infty$ for $\tail_P \in \mcalr_0$. 
	For case (2), by \thmref{thm:Yinfty}, we have $\sinr \overset{d}{\to} D$, where $D$ has a non-degenerate distribution. Then, for constant $u > 0$ and $y > 0$, $\Pb(\sinr(u\lambda) \geq y) = \Pb(D \geq y) = \Pb(\sinr(\lambda) \geq y)$ as $\lambda \to \infty$. Similarly, $\E\pren{\log(1 + \sinr(u\lambda))} = \E\pren{\log(1+D)} = \E\pren{\log(1 + \sinr(\lambda))}$ as $\lambda \to \infty$.
	
	For case (3), given the conditions and \thmref{thm:Yinfty}, we have $\sinr \overset{a.s.}{\to} 0$. Hence, $\lim_{\lambda \to \infty}\Pp_y(\lambda) = 0$ and by \lemref{lem:Yzero}, $\exists\lambda_p > 0$ s.t. $\Pp_y(\lambda_p) > \Pp_y(0) = 0$. For $\Cc(\lambda)$, we first note that $\Cc(\lambda) = \int_{0}^{\infty}\Pb(\sinr(\lambda) > y)/(1+y)\d{y}$. Thus, $\lim_{\lambda \to \infty}\Cc(\lambda) = 0$ by Lebesgue's dominated convergence theorem and by the fact that $\sinr \overset{p}{\to} 0$. For $\epsilon > 0$, $\exists\lambda_c > 0$ such that $\Pb(\sinr(\lambda_c) \geq \epsilon) > \Pb(\sinr(0) \geq \epsilon) = 0$ due to \lemref{lem:Yzero}. Thus, $\forall y \in [0, \epsilon]$, $\Pb(\sinr(\lambda_c) \geq y) > 0$ since $\Pb(\sinr \geq y)$ is decreasing w.r.t. $y$. In consequence, $\Cc(\lambda_c) = \int_{0}^{\infty}\Pb(\sinr(\lambda_c) > y)/(1+y)\d{y} > 0$.

\subsection{Proof of \corref{cor:NwkLimit}}\label{appen_corNwkLimit}
	By noting that $\Dc_y = \lambda\Pp{y}$ and $\Ac = \lambda \Cc$ due to \eref{eq:PerfDefn} and \eref{eq:PerfDefn2}, the first and the third assertions directly follow from \thmref{thm:PerfLimit}.
	
	For the second assertion, we have $\sinr \tod D$ where $D$ has a non-degenerate distribution depending on the regularly varying index $\alpha$ of $\tail_P$ due to \thmref{thm:Yinfty}. Thus, $\Dc_y(\lambda) = c_{y,\alpha}\lambda$ as $\lambda \to \infty$ in which constant $c_{y,\alpha} := \Pb(D \geq y)$ depends on $y$ and $\alpha$. Similarly, $\Ac(\lambda) = \lambda \E(\log(1 + D))$ as $\lambda \to \infty$. Since $D$ has a non-degenerate distribution and $D \geq 0$, we have $c_{\alpha} := \E(\log(1+D)) > 0$ and $c_{\alpha}$ depends on $\alpha$. Hence the proof.

\subsection{Proof of \lemref{lem:LaplaceZ}}\label{appen_Lemma4}
	For positive integer $n$, define $S_n = \sum_{i=1}^n P_i$, $M_n = \max_{i=1}^n P_i$, and put $Z_n = (S_n - M_n)/M_n$. We first derive the Laplace transform of $Z_n$ using the same technique exposed in \cite[Lemma 2.1]{Darling1952}. Without lost of generality assume that $M_n = P_1$, which has probability $1/n$. Then, for $s \in \Cb$ and $\Re(s) \geq 0$,
	\begin{multline*}
	\Lc_{Z_n}(s) = \E(e^{-s Z_n}) \\
	= n\int_0^{\infty}\cdots\int e^{-s(x_2 + \cdots + x_n)/x_1}g(x_1,\cdots,x_n)\d{x_1}\cdots\d{x_n}
	\end{multline*}
	where $g(x_1,\cdots,x_n)$ is the joint density of $X_1, \cdots, X_n$ given that $X_1 = M_n$:
	\begin{equation*}
	g(x_1,\cdots,x_n) = \begin{cases}
	f_P(x_1)\cdots f_P(x_n) & \text{if } x_1 = \max_{i=1}^n x_i \\
	0 & \text{otherwise}
	\end{cases}.
	\end{equation*}
	Thus
	\begin{align}
	& \Lc_{Z_n}(s) % = n \int_0^{\infty} \int_0^{x} \cdots \int_0^x e^{-s(x_2 + \cdots + x_n)/x}f_P(x_2)\cdots f_P(x_n)f_P(x)\d{x_2}\cdots\d{x_n}\d{x} \nonumber\\
	= n \int_0^{\infty} \int_0^{x} \cdots \int_0^x \pren{\prod_{i=2}^n{e^{-s \frac{x_i}{x}}f_P(x_i)\d{x_i}}} f_P(x)\d{x} \nonumber\\
	& = n \int_0^{\infty} \left( x\int_0^{1} e^{-st}f_P(xt)\d{t}\right)^{n-1} f_P(x)\d{x} \nonumber \\
	& = n \int_0^{\infty} \left(\varphi(x)\right)^{n-1} f_P(x)\d{x}, \label{eq:LaplaceZn}
	\end{align}
	where $\varphi(x) \defeq x\int_0^{1} e^{-st}f_P(xt)\d{t}$. Here, firstly we note that
	\begin{equation*}
	|\varphi(x)| \leq \int_0^1 |x e^{-st} f_P(xt)|\d{t} = \int_0^1 x f_P(xt)\d{t} = F_P(x) < 1,
	\end{equation*}
	for $x < \infty$. Thus, $n \int_0^{T} \left(\varphi(x)\right)^{n-1} f_P(x)\d{x} \to 0$ as $n \to \infty$ for any $T < \infty$. As a consequence, we only need to consider the contribution of large $x$ in \eqref{eq:LaplaceZn} for $\Lc_{Z_n}(s)$. It follows that, an integration by parts with $e^{-st}$ and $xf_P(xt)\d{t}$ yields
	\begin{align*}
	\varphi(x) & = 1 - e^{-s}\tail_P(x) - \int_0^1 s e^{-st} \tail_P(xt) \d{t} \\
	& = 1 - \tail_P(x) + \int_0^1 s e^{-st} \left(\tail_P(x) - \tail_P(tx)\right)\d{t}.
	\end{align*}
	For $\tail_P \in \mcalr_{-\alpha}$ with $\alpha \in [0,\infty)$, by representation theorem, see e.g. \cite[Theorem 1.4.1]{Bingham1989}, we can write $\tail_P(tx) \sim t^{-\alpha}\tail_P(x)$ for $0 < t < \infty$ and $x \to \infty$. Thus
	\begin{multline*}
	\int_0^1 s e^{-st} \left(\tail_P(x) - \tail_P(tx)\right) \d{t} \\
	\sim \tail_P(x)\int_0^1 s e^{-st} (1 - t^{-\alpha})\d{t} \\
	= -\tail_P(x) \int_0^1 \alpha (1 - e^{-st}) \frac{\d{t}}{t^{\alpha + 1}}
	\end{multline*}
	for large $x$, hence
	\begin{equation*}
	\varphi(x) = 1 - (1 + \phi)\tail_P(x), \quad \text{for } x \to \infty,
	\end{equation*}
	where $\phi \defeq \int_0^1 \alpha (1 - e^{-st}) \frac{\d{t}}{t^{\alpha + 1}}$. Substitute $\varphi(x)$ back to the expression of $\Lc_{Z_n}(s)$ we obtain
	\begin{equation*}
	\Lc_{Z_n}(s) \sim n \int_0^{\infty} \left( 1 - (1 + \phi)\tail_P(x) \right)^{n-1} f_P(x)\d{x},
	\end{equation*}
	as $n \to \infty$. Here, use a change of variable with $v = n \tail_P(x)$, we have
	\begin{align*}
	\Lc_{Z_n}(s) & \sim \int_0^n \left(1 - \frac{v}{n}(1+\phi)\right)^{n-1}\d{v} \\
	& \overset{(a)}{\to} \int_0^\infty e^{-v(1+\phi)}\d{v} = (1 + \phi)^{-1}, \quad \text{ as } n \to \infty,
	\end{align*}
	where notice that $(a)$ is due to $(1 + \frac{x}{n})^n \to e^x$ as $n \to \infty$. Finally, since noise power $W < \infty$ and $M \overset{a.s.}{\to} \infty$ as $\lambda \to \infty$, we have $\lim_{\lambda \to \infty } Z = \lim_{n \to \infty} Z_n$. Hence the proof. %\lemref{lem:LaplaceZ} for $\alpha \in [0,\infty)$.

\subsection{Proof of \thmref{thm:Ordering}}\label{appen_Theorem4}
	Firstly, since $\alpha/t^{\alpha+1}$ is increasing with respect to $\alpha \geq 0$ for $t \in [0,1]$, $\phi$ is increasing  w.r.t. $\alpha \geq 0$. Thus, using \lemref{lem:LaplaceZ}, $\Lc_Z(s)$ with $s > 0$ is decreasing w.r.t. $\alpha \in [0,\infty)$ as $\lambda \to \infty$. Hence, by using the same notation of \lemref{lem:LaplaceZ} and let $Z_1 = 1/\sinr_1$ and $Z_2 = 1/\sinr_2$, we have, $Z_{1} \leq_{\rm Lt} Z_{2}$ for $0 \leq \alpha_1 \leq \alpha_2 < +\infty$. Now, for $\alpha_2 = +\infty$, we have $\sinr_2 \overset{a.s.}{\to} 0$ by \thmref{thm:Yinfty}, thus $Z_2 \overset{a.s.}{\to} +\infty$, resulting in $\Lc_{Z_2}(s) = 0$, $\forall s > 0$. Therefore, the Laplace ordering $Z_1 \leq_{\rm Lt} Z_2$ is still verified.
	
	This also due to the fact that since $\displaystyle \lim_{x\to\infty}\pren{\bar{F}_{Z_1}(x)/\bar{F}_{Z_2}(x)} = 0$, there exists $x_0 > 0$ such that, for any $x \geq x_0$, $\bar{F}_{Z_1}(x)<\bar{F}_{Z_2}(x)$, hence $\bar{F}_{Z_1}(x) \leq \bar{F}_{Z_2}(x)$, $\forall x \in \mathbb{R}^+$ is satisfied at infinity, i.e. $Z_2$ strictly dominates $Z_1$. As $Z_1 \leq_{\mathrm{st}} Z_2$ and $g(x) = e^{-sx}$ is c.m. function for $s\geq0$, we have $Z_1 \leq_{\mathrm{Lt}} Z_2$. Since $g(x) = 1/x$ is a c.m. function, we have that $\mathbb{E}(\sinr_1) \geq \mathbb{E}(\sinr_1)$. Finally, the function $g(x) = \log(1+\frac{1}{x})$ is also c.m. function, hence from stochastic ordering results, we have that $\mathbb{E}(\log(1+\sinr_1)) \geq \mathbb{E}(\log(1+\sinr_2))$.  

%%%%%%%%%%%%%%%%%%%%%%%%%%%%%%%%%%%%%%%%%%%%%%%%%%%%%%%%%%%%%%%%%%

\bibliographystyle{IEEEtran}
\bibliography{IEEEabrv,Densification}

\begin{IEEEbiography}%[{\includegraphics[width=1in,height=1.25in,clip,keepaspectratio]{Nguyen.png}}]
	{Van Minh Nguyen} received the M.Eng and M.Sc degrees in Electrical and Computer Engineering from Telecom Bretagne in 2007, the Certificate in management from ENPC MBA Paris in 2009, and the PhD degree in Electrical Engineering from Telecom ParisTech in 2011. His PhD thesis, funded by Alcatel-Lucent Bell Laboratories, was realized at Bell Laboratories France and Network Theory and Communications (TREC) research lab of INRIA-École Normale Sup\'{e}rieure.

	From 2005 to 2010, he was with Alcatel (then Alcatel-Lucent) as an Engineer during 2005-2006, a Research Intern during Mar. - Oct. 2007, and a Research Engineer at Bell Laboratories from Nov. 2007 to Oct. 2010. From Nov. 2010 to Mar. 2011, he was a Research Engineer at INRIA, Paris, France. From Mar. 2011 to Dec. 2014, he was with the R\&D department of Sequans Communications, France, as a Research Scientist - DSP Engineer working on algorithm development for LTE terminal modem. Since Dec. 2014, he has been a Senior Research Scientist at the Mathematical and Algorithmic Sciences Laboratory, Huawei Technologies, France. His research includes wireless communications, digital signal processing, coding, wireless network modeling, and radio resource management. He is the author of three granted patents, three contributions to 3GPP standard, and several patent applications. He received the Award of the French Embassy in Vietnam, the Award of Nortel, the Award of Alcatel for the best Engineer of the year, and several internal awards from Sequans Communications and Huawei Technologies.	
\end{IEEEbiography}

\begin{IEEEbiography}%[{\includegraphics[width=1in,height=1.25in,clip,keepaspectratio]{Kountouris.jpg}}]
	{Marios Kountouris} (S'04–-M'08–-SM'15) received the Diploma in Electrical and Computer Engineering from the National Technical University of Athens, Greece in 2002 and the M.S. and Ph.D. degrees in Electrical Engineering from the Ecole Nationale Supérieure des Télécommunications (Télécom ParisTech), France in 2004 and 2008, respectively. His doctoral research was carried out at Eurecom Institute, France, and was funded by Orange Labs, France. From February 2008 to May 2009, he has been with the Department of ECE at The University of Texas at Austin as a research associate, working on wireless ad hoc networks under DARPA’s IT-MANET program. From June 2009 to December 2013, he has been an Assistant Professor at the Department of Telecommunications at SUPELEC (Ecole Supérieure d’Electricité – now CentraleSupélec), France, and from January 2014 to July 2016, he has been an Associate Professor. From March 2014 to February 2015, he has been an Adjunct Professor in the School of EEE at Yonsei University, S. Korea. Since January 2015, he has been a Principal Researcher at the Mathematical and Algorithmic Sciences Lab, Huawei Technologies, France. 
	
	He currently serves as Associate Editor for the IEEE Transactions on Wireless Communications, the IEEE Transactions on Signal Processing, the IEEE Wireless Communication Letters, and Networking, and the Journal of Communications and Networks (JCN). He received the 2016 IEEE ComSoc Communication Theory Technical Committee Early Achievement Award, the 2013 IEEE ComSoc Outstanding Young Researcher Award for the EMEA Region, the 2014 Best Paper Award for EURASIP Journal on Advances in Signal Processing (JASP), the 2012 IEEE SPS Signal Processing Magazine Award, the IEEE SPAWC 2013 Best Student Paper Award, and the Best Paper Award in Communication Theory Symposium at IEEE Globecom 2009. He is a Professional Engineer of the Technical Chamber of Greece.	
\end{IEEEbiography}
\vfill
\end{document}